\documentclass[journal]{IEEEtran}
\usepackage{amsfonts}
\usepackage{amssymb}
\usepackage{url}
\usepackage{array}
\usepackage{multirow,multicol}
\usepackage{threeparttable}
\usepackage{tablefootnote}
\usepackage[super]{nth}
\usepackage{cite}
\usepackage{makecell}
\usepackage{booktabs}
\usepackage[linesnumbered, ruled]{algorithm2e}
\usepackage{amsmath}
\usepackage{xcolor}
\usepackage{graphicx}
\usepackage{mathrsfs}
\usepackage{tabularx}
\usepackage{verbatim}

\ifCLASSINFOpdf
\else
\fi
\begin{document}
%
\title{Online Hybrid CTC/attention End-to-End Automatic Speech Recognition Architecture}
%
%
%

\author{Haoran Miao,~\IEEEmembership{Student Member,~IEEE,}
        Gaofeng Cheng,~\IEEEmembership{Member,~IEEE,}
        Pengyuan Zhang,~\IEEEmembership{Member,~IEEE,}\\
        Yonghong Yan,~\IEEEmembership{Member,~IEEE}
\thanks{Manuscript received xxxxxxxx; revised xxxxxxxx. (Corresponding author: Gaofeng Cheng.)}
\thanks{H. Miao (e-mail: miaohaoran@hccl.ioa.ac.cn), G. Cheng (e-mail: chenggaofeng@hccl.ioa.ac.cn), P. Zhang (e-mail: zhangpengyuan@hccl.ioa.ac.cn) and Y. Yan (e-mail: yanyonghong@hccl.ioa.ac.cn) are with Institute of Acoustics, Chinese Academy of Sciences and University of Chinese Academy of Sciences, Beijing, China. }
}

\maketitle

\begin{abstract}Recently, there has been increasing progress in end-to-end automatic speech recognition (ASR) architecture, which transcribes speech to text without any pre-trained alignments. One popular end-to-end approach is the hybrid Connectionist Temporal Classification (CTC) and attention (CTC/attention) based ASR architecture, which utilizes the advantages of both CTC and attention. The hybrid CTC/attention ASR systems exhibit performance comparable to that of the conventional deep neural network (DNN) / hidden Markov model (HMM) ASR systems. However, how to deploy hybrid CTC/attention systems for online speech recognition is still a non-trivial problem. This paper describes our proposed online hybrid CTC/attention end-to-end ASR architecture, which replaces all the offline components of conventional CTC/attention ASR architecture with their corresponding streaming components. Firstly, we propose stable monotonic chunk-wise attention (sMoChA) to stream the conventional global attention, and further propose monotonic truncated attention (MTA) to simplify sMoChA and solve the training-and-decoding mismatch problem of sMoChA. Secondly, we propose truncated CTC (T-CTC) prefix score to stream CTC prefix score calculation. Thirdly, we design dynamic waiting joint decoding (DWJD) algorithm to dynamically collect the predictions of CTC and attention in an online manner. Finally, we use latency-controlled bidirectional long short-term memory (LC-BLSTM) to stream the widely-used offline bidirectional encoder network. Experiments with LibriSpeech English and HKUST Mandarin tasks demonstrate that, compared with the offline CTC/attention model, our proposed online CTC/attention model improves the real time factor in human-computer interaction services and maintains its performance with moderate degradation. To the best of our knowledge, this is the first work to provide the full-scale online solution for CTC/attention end-to-end ASR architecture. \par

\end{abstract}


\IEEEpeerreviewmaketitle

\section{Introduction}

\IEEEPARstart{I}{n} the past decade, we have witnessed impressive progress in automatic speech recognition (ASR) field. The hybrid deep neural network (DNN)/hidden Markov model (HMM) \cite{dahl2011context, hinton2012deep, vesely2013sequence} is the first successful deep-learning ASR architecture. A typical hybrid DNN/HMM system is composed of several separate modules, including acoustic, pronunciation and language models (AM, PM, LM), all of which are designed manually and optimized separately. Additionally, the hybrid DNN/HMM system needs a complex decoder which has to be performed by integrating AM, PM and LM. To conclude, the hybrid DNN/HMM ASR models require linguistic information and complex decoders, and thus it is difficult to develop hybrid DNN/HMM ASR systems for new languages.

In recent years, end-to-end speech recognition \cite{graves2014towards,CTC_Graves_2006,watanabe2017hybrid,seq2seq_state_of_the_art_2018_google,chorowski2015attention,chan2016listen} has gained popularity in the ASR community.
End-to-end ASR models simplify the conventional DNN/HMM ASR system into a single deep neural network architecture. Besides, the end-to-end models require no lexicons and predict graphemes or words directly, which makes the decoding procedure much simpler than those of the hybird DNN/HMM models. 
To date, the end-to-end ASR architectures have gained significant improvement in speech recognition accuracy \cite{amodei2016deep, watanabe2017hybrid, seq2seq_state_of_the_art_2018_google}. There are two major types of end-to-end ASR architectures. One is Connectionist Temporal Classification (CTC) based ASR architecture \cite{CTC_Graves_2006,jinyu_ctc_word_2018,jinyu_ctc_trans_2019} and the other is the attention-based ASR architecture \cite{bahdanau2014neural,chorowski2015attention,chan2016listen, seq2seq_state_of_the_art_2018_google}. 
To combine theses two architectures, the hybrid CTC/attention architecture \cite{kim2017joint, watanabe2017hybrid} was proposed to leverage the CTC objective as an auxiliary task in an attention-based encoder-decoder network. During training, a CTC objective is attached to an attention-based encoder-decoder model as a regularization.
During decoding, a joint CTC/attention decoding approach was proposed to combine the decoder scores and CTC scores in the beam search algorithm. Specifically, the CTC branch uses the CTC prefix score \cite{kawakami2008supervised} to efficiently exclude hypotheses with poor alignments. By taking the advantages of both CTC and attention mechanisms, the hybrid CTC/attention architecture has been proved to be superior to simple attention-based and CTC-based models \cite{watanabe2017hybrid}. \par

Although the hybrid CTC/attention end-to-end ASR architecture is reaching reasonable performance \cite{karita2019transformer,Wang2019,das2019ctcatt,haoran2020online}, how to deploy it in online scenarios remains an unsolved problem. After inspections of the CTC/attention ASR architecture, we identify four challenges to deploy online hybrid CTC/attention end-to-end ASR systems:\par
\begin{itemize}
    \item \textbf{Attention mechanism}: The conventional CTC/attention ASR architecture employs global attention mechanism like additive attention \cite{bahdanau2014neural} or location-aware attention \cite{chorowski2015attention}, which performs attention over the entire input representations. 
    \item \textbf{CTC prefix score}: CTC prefix score is defined as the cumulative probability of all label sequences sharing the same prefix. We need the information of complete utterances to compute the CTC prefix scores.
    \item \textbf{Unsynchronized predictions between CTC and attention}: 
    Attention-based ASR models perform label synchronous decoding, while CTC-based ASR models are frame-synchronized. Even though CTC/attention joint training can guide them to have more synchronized predictions, the predictions of CTC and attention are not strictly synchronized. This phenomenon makes online joint CTC/attention decoding difficult.
    \item \textbf{Bidirectional encoder}: Bidirectional long short-term memory (BLSTM) \cite{Hochreiter1997Long,Schuster1997Bidirectional} networks, which can exploit long-term dependency of input speeches, are widely used in CTC/attention end-to-end ASR architectures. However, the bidirectional encoder hampers the online deployment of hybrid CTC/attention models.
\end{itemize}

To surmount the obstacles of deploying online attention based end-to-end models, there have been some prior published efforts to stream attention mechanisms, including neural transducer (NT) \cite{jaitly2016online, sainath2018improving}, hard monotonic attention (HMA) \cite{raffel2017online}, monotonic chunk-wise attention (MoChA) \cite{chiu2017monotonic}, triggered attention (TA) \cite{niko2019}, etc. NT is a limited-sequence streaming attention-based model that consumes a fixed number of input frames and produces a variable number of labels before the next chunk of input frames arrive. However, NT requires coarse alignments during training (e.g., what words are verbalized by what chunk of input frames). HMA and MoChA are alignment-free streaming attention models, which enables the attention to automatically select a fixed number of input representations in monotonic left-to-right mode. However, \cite{online-trans} found that MoChA had convergence problem because the attention weights of MoChA decreased immediately, and \cite{google-long-utt} found that MoChA had problems scaling up to long utterances. TA was proposed to leverage CTC-based networks to dynamically partition utterances, and to perform attention on the incremental input representations as the decoder generates labels. \par

\begin{figure}[t]
    \centering
    \includegraphics[scale=0.43]{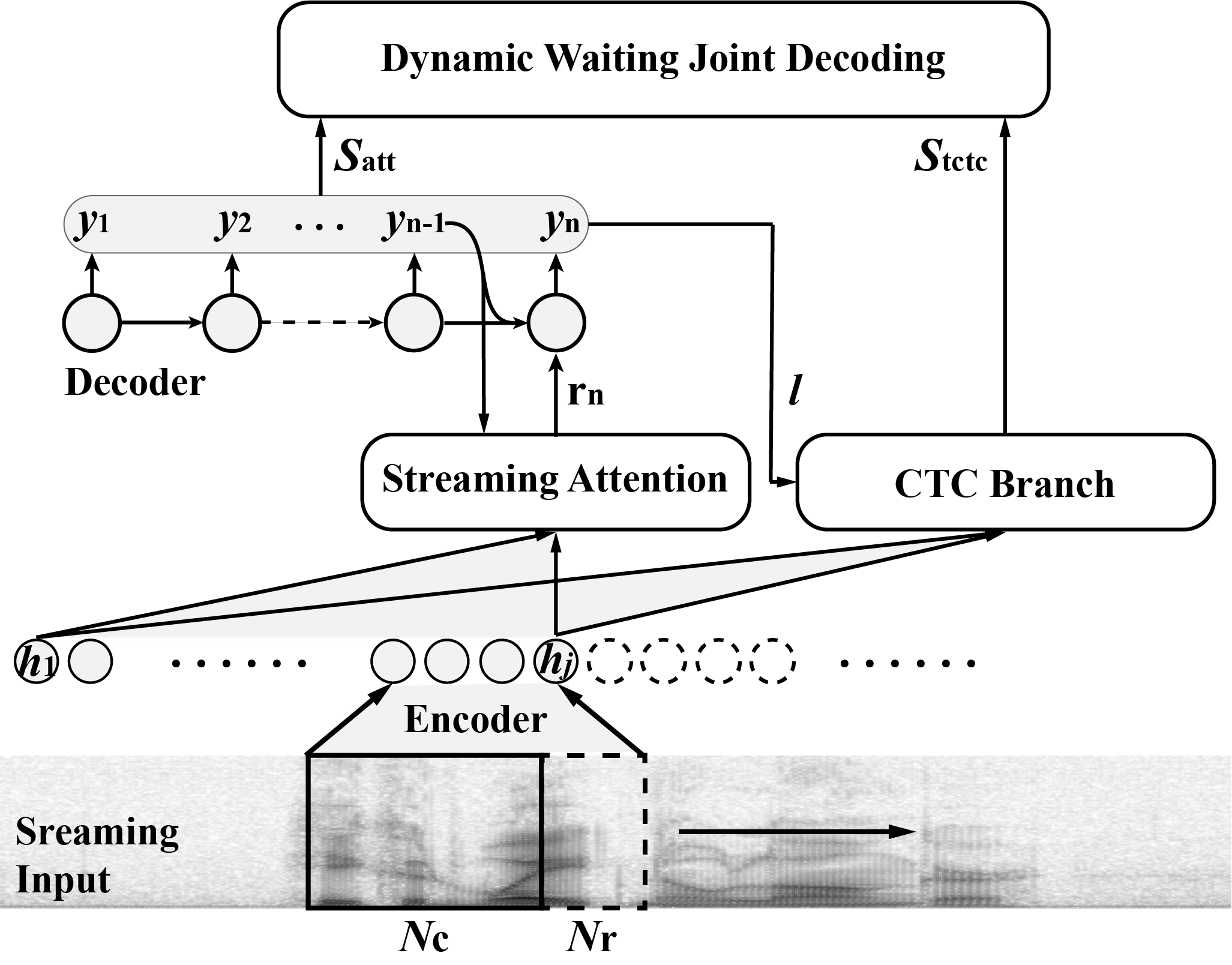}
    \caption{The proposed online hybrid CTC/attention based end-to-end ASR architecture. The architecture uses streaming attention mechanism to stream the attention branch, and uses DWJD algorithm to dynamically collect decoding scores from attention-based decoder ($S_{\rm{att}}$) and CTC branch ($S_{\rm{tctc}}$).}
    \label{fig:structure}
\end{figure}

Suffering from the joint CTC/attention decoding strategy, it requires sophisticated skills to realize the online CTC/attention end-to-end ASR architecture. This paper summarizes our work on streaming the CTC/attention ASR architecture in four aspects.
First, we propose a stable MoChA (sMoChA) to address the non-convergence problem of MoChA. Furthermore, we design a monotonic truncated attention (MTA) to deal with the training-and-decoding mismatch problem of sMoChA. Compared with sMoChA, MTA is more simple and achieves higher recognition accuracy. Second, we leverage CTC output to segment input representations and compute a truncated CTC (T-CTC) prefix score on top of the segmented input representations rather than the complete utterances. Our experiments demonstrate that the T-CTC prefix score can approximate to its corresponding CTC prefix score and accelerate the decoding speed. Third, we propose a dynamic waiting joint decoding (DWJD) algorithm to solve the unsynchronized predictions problem of joint CTC/attention decoding. Finally, we use a stack of VGGNet-style \cite{simonyan2014very} convolutional neural networks (CNNs) as the encoder front-end and apply latency-controlled BLSTM (LC-BLSTM) \cite{zhang2016highway,jain2019lcblstm} as the encoder back-end. Figure~\ref{fig:structure} depicts our proposed online CTC/attention end-to-end ASR architecture. \par

Our experiments showed that almost no speech recognition accuracy degradation was caused by using MTA, T-CTC and DWJD, which demonstrated the superiority and robustness of our proposed methods. The major recognition accuracy degradation came from the using of LC-BLSTM based encoder networks. \par

Compared with our preliminary version\cite{haoran2019online}, the new content in this paper includes proposing MTA to improve the online performance, providing more details of decoding algorithms and verifying the validity of the system in Chinese and English by more experiments. 
The rest of this paper is organized as follows. In section \uppercase\expandafter{\romannumeral2}, we describe the hybrid CTC/attention ASR architecture. In section \uppercase\expandafter{\romannumeral3}, we present some prior streaming attention methods. In section \uppercase\expandafter{\romannumeral4}, we describe the proposed online CTC/attention end-to-end ASR architecture. The experimental setups and results are described in section \uppercase\expandafter{\romannumeral5}. Finally we make the conclusions in section \uppercase\expandafter{\romannumeral6}. \par

\section{Hybrid CTC/attention Architecture}
In this section, we introduce the training and decoding details about the hybrid CTC/attention end-to-end ASR architecture \cite{watanabe2017hybrid,CTC-Attention-ACL-2017} and reveal why it is difficult to deploy it in online scenarios. \par

\subsection{Training of Hybrid CTC/attention ASR Architecture}
The hybrid CTC/attention end-to-end ASR architecture is based on the attention-based encoder-decoder model \cite{chorowski2015attention}. Suppose the $T$ input frames $X=(x_1, \cdot\cdot\cdot, x_T)$ correspond to the $L$ output labels $Y=(y_{1}, \cdot\cdot\cdot, y_{L})$, the encoder converts $X$ into input representation vectors $H=(\mathbf{h}_1, \cdot\cdot\cdot, \mathbf{h}_T)$. To create a context for each representation vector that depends on both its past as well as its future, we usually choose the stack of BLSTM as the offline encoder network. Besides, there are also alternatives for low-latency encoders, such as LSTM, LC-BLSTM and etc. \par

The attention-based decoder is an autoregressive network and computes the conditional probabilities for each label $y_i$ as follows:
\begin{eqnarray}
    \mathbf{a}_{i}&=&{\rm GlobalAttention}(\mathbf{a}_{i-1}, \mathbf{q}_{i-1}, H),\label{convec}\\
		\mathbf{r}_i&=&\sum_{j=1}^{T} a_{i,j} \mathbf{h}_j, \label{att_sum}
\end{eqnarray}
\begin{equation}
     p(y_{i}|y_{1}, \cdot\cdot\cdot, y_{i-1}, X)={\rm{Decoder}}(\mathbf{r}_i,\ \mathbf{q}_{i-1},\ y_{i-1}), \label{decy} 
\end{equation}
where $i$ and $j$ are indices of output labels and input representation vectors, respectively, $\mathbf{a}_{i}\in\mathbb{R}^T$ is a vector of the attention weights, $\mathbf{q}_{i-1}$ is the $(i-1)$-th state of the decoder, $\mathbf{r}_i$ is a label-wise representation vector. \par

In the state-of-the-art end-to-end ASR models \cite{chorowski2015attention,Park2019}, the global attention is implemented as the location-aware attention (LoAA). For $j=1,\cdot\cdot\cdot,T$, LoAA computes the attention weights as follows:
\begin{eqnarray}
	\mathbf{f}_{i}&=&\mathbf{Q}\ast \mathbf{a}_{i-1},\label{loaa1}\\
	e_{i,j}&=&\mathbf{v}^\top{\rm{tanh}}(\mathbf{W}_{1}\mathbf{q}_{i-1}+\mathbf{W}_{2}\mathbf{h}_{j}+\mathbf{W}_{3}\mathbf{f}_{i}+\mathbf{b}),\label{loaa2} \\
	a_{i,j}&=&{\rm{Softmax}}(\{e_{i,j}\}_{j=1}^{T}),\label{loaa3}
\end{eqnarray}
where matrices $\mathbf{W}_{1}$, $\mathbf{W}_{2}$, $\mathbf{W}_{3}$, $\mathbf{Q}$ and vectors $\mathbf{b}$, $\mathbf{v}$ are learnable parameters. The term $\ast$ denotes an one-dimensional convolution operation, with the convolutional parameter, $\mathbf{Q}$, along the input frame axis $j$. The computational complexity of LoAA is $\mathcal{O}(TL)$. \par

Different from the basic encoder-decoder model, the hybrid CTC/attention end-to-end ASR architecture leverages the CTC \cite{CTC_Graves_2006} objective as an auxiliary task during training. Specifically, the CTC branch shares the encoder network and has an additional classification layer to predict labels or ``blank'' label $\langle b\rangle$. Therefore, the objective function $\mathcal{L}_{mtl}$ is made up of a linear interpolation of the CTC and attention objectives, which is expressed as 
\begin{equation}
    \mathcal{L}_{mtl}=\lambda \mathcal{L}_{\rm{ctc}} + (1\!-\!\lambda)\mathcal{L}_{\rm{att}},
    \label{mot}
\end{equation}
where the tunable coefficient $\lambda$ satisfies $0\!\leq\!\lambda\!\leq\!1$.

\subsection{Decoding of Hybrid CTC/attention ASR Architecture}
The previous work \cite{watanabe2017hybrid} shows that the attention-based encoder-decoder model tends to perform poor alignment without additional constraints, prompting the decoder to generate unnecessarily long hypotheses. In the hybrid CTC/attention architecture, the joint CTC/attention decoding method helps to refine the search space \cite{watanabe2017hybrid,CTC-Attention-ACL-2017} by considering the CTC prefix scores \cite{kawakami2008supervised} of hypotheses. \par

However, the joint decoding method is inapplicable in online scenarios not only for the global attention, but also for the CTC prefix scores.
Suppose the decoder generate a partial hypothesis $l=(y_{1}, \cdot\cdot\cdot, y_{n})$, where $n$ is the length of this hypothesis, the score of $l$ assigned by the decoder branch is
\begin{equation}
    S_{\rm{att}}=\sum_{i=1}^{n}{\rm{log}}\ p_{\rm{att}}(y_{i}|y_{1},\cdot\cdot\cdot,y_{i\!-\!1},H).
\end{equation}
On the one hand, because the global attention aligns each output label to the entire input representation vectors, the attention-based decoder depends on the full utterance; on the other hand, because the attention provides no information about the time boundary of $l$, it is difficult for the CTC branch to compute the exact probability of $l$. Therefore, in the joint decoding method, the CTC branch leverages the CTC prefix score to consider all possible time boundaries, which requires the full utterance. The score of $l$ assigned by the CTC branch is
\begin{equation}
    S_{\rm{ctc}}={\rm{log}}\sum_{j=1}^{T}p_{\rm{ctc}}(l|H_{1:j}).\label{ctcsum}
\end{equation}

To further improve the decoding accuracy, we employ an external RNN language model \cite{hori2017advances}, which is separately trained with the training transcriptions or extra corpus. In the joint decoding framework, a beam search is applied to prune $l$ in accordance with the scoring function
\begin{equation}
    S=\mu S_{\rm{ctc}} + (1\!-\!\mu)S_{\rm{att}} + \beta S_{\rm{lm}},
\end{equation}
where $S_{\rm{lm}}$ is the score from the language model, $\mu$ and $\beta$ represent the CTC and language model weights, respectively adjusting the proportion of different scores in the beam search. For the decoding end detection criteria, if there is little chance of finding a longer complete hypothesis with higher score, the beam search will be terminated \cite{watanabe2017hybrid}. \par

\section{Prior Streaming Attention Works}
The attention-based end-to-end models typically leverage a global attention network to perform alignments. To apply online CTC/attention decoding, we first need to stream the attention mechanisms. In this section, we introduce some recent prior efforts on streaming the attention mechanisms. \par

\begin{figure}
    \centering
    \includegraphics[scale=0.65]{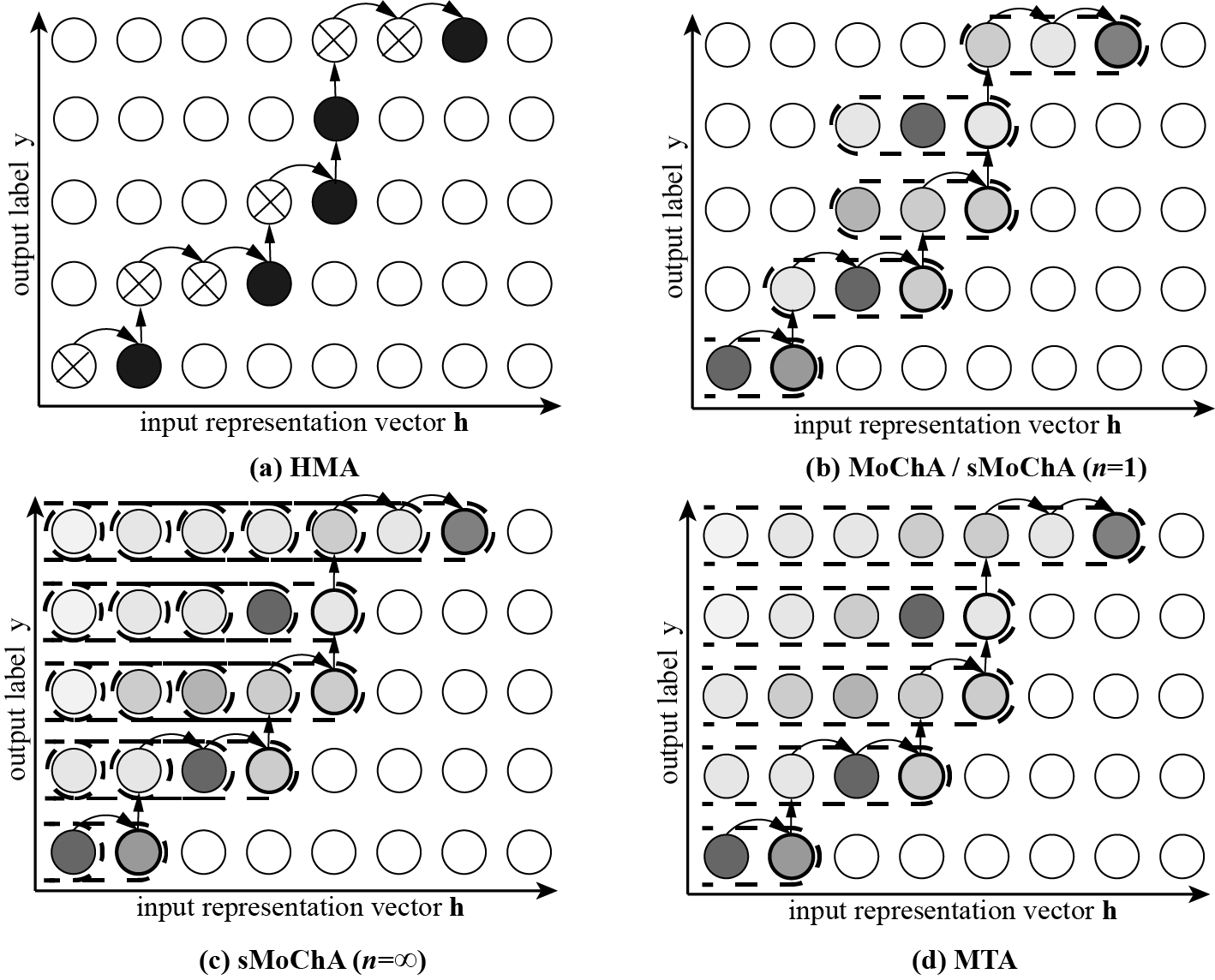}
    \caption{Schematics of various streaming attention mechanisms in the decoding stage. End-points are shown as black nodes in (a) and nodes with bold borders in (b)-(d). Curved arrow indicates how these end-points move at each output step. Dotted lines in (b)-(d) indicate which input representation vectors align to attention models.}
    \label{fig:att}
\end{figure}

\subsection{Hard Monotonic Attention (HMA)}
In many ASR tasks, according to the observations \cite{chorowski2015attention, chan2016listen}, the attention tends to align each output label to the input representations nearly locally and monotonically \cite{chorowski2015attention, chan2016listen}. Based on this property, HMA aligns each output label to a single input representation in a monotonic left-to-right way, as shown in figure~\ref{fig:att}(a).
\subsubsection{Decoding of HMA}
In the decoding stage, HMA always begins a search from a previously attended input representation and chooses the next input representation via a stochastic process. Let $i$ and $j$ denote the indices of the output labels and input representation vectors, respectively. Assuming that HMA has attended to the $t_{i\!-\!1}$-th input representation vector $\mathbf  {h}_{t_{i\!-\!1}}$ at the ($i\!-\!1$)-th output time-step, the stochastic process selects the next input representation vector by a sequential method, e.g., for $j=t_{i\!-\!1},\ t_{i\!-\!1}\!+\!1,\ ...,$
\begin{eqnarray}
    e_{i,j}&=&{\rm{Energy}}(\mathbf{q}_{i-1},\mathbf{h}_{j}),\label{mhae}\\
    p_{i,j}&=&{\rm{Sigmoid}}(e_{i,j}),\\
    z_{i,j}&\sim&{\rm{Bernoulli}}(p_{i,j})\label{mhaz},
\end{eqnarray}
where $\mathbf{q}_{i-1}$ is the $(i-1)$-th state of the decoder, $p_{i,j}$ is a selection probability and $z_{i,j}$ is a discrete variable. Once we sample $z_{i,j}=1$ for some $j$, HMA stops the searching process and sets the label-wise representation vector $\mathbf{r}_{i}=\mathbf{h}_{j}$. To make the decoding process deterministic and efficient, \cite{raffel2017online} replaced Bernoulli sampling with an indicator function $z_{i,j}=\mathbb{I}(p_{i.j}>0.5)$. \par

The decoding computational complexity of HMA is $\mathcal{O}(T)$, where $T$ is the length of the sequential input representation, because HMA scans each input representation vector only once. \par 

\subsubsection{Training of HMA}
As the discrete variable $z_{i,j}$ is conflict to backpropagation algorithm in the training stage, we have to compute the expectations $\mathbb{E}[z_{i,j}]$ and $ \mathbb{E}[\mathbf{r}_{i}]$ based on all the input representation vectors during training:
\begin{eqnarray}
    \mathbb{E}[z_{i,j}]&=&p_{i,j}\sum_{k=1}^{j}\left(\mathbb{E}[z_{i\!-\!1,k}]\prod_{l=k}^{l\!-\!1}(1-p_{i,l})\right),\label{mha}\\
    \mathbb{E}[\mathbf{r}_{i}]&=&\sum_{j=1}^{T}\mathbb{E}[z_{i,j}]\cdot \mathbf{h}_{j}.
\end{eqnarray}
To analyze the convergence property of $\mathbb{E}[z_{i,j}]$. Equation (\ref{mha}) can be rewritten in recursion:
\begin{equation}
    \mathbb{E}[z_{i,j}]=\frac{p_{i,j}}{p_{i,j\!-\!1}}\cdot(1-p_{i,j\!-\!1})\mathbb{E}[z_{i,j\!-\!1}]+p_{i,j}\mathbb{E}[z_{i\!-\!1,j}].\label{conva}
\end{equation}
According to the term $\frac{p_{i,j}}{p_{i,j\!-\!1}}\cdot(1-p_{i,j\!-\!1})\mathbb{E}[z_{i,j\!-\!1}]$, the expectations of $z$ exponentially decay by $(1-p_{i,j\!-\!1})$ along the index $j$ \cite{raffel2017online}, which means that it is difficult for the attention network to attend to the input representation vector at long distance.
Therefore, \cite{raffel2017online} used a modified \emph{Energy} function as follows:
\begin{equation}
    e_{i,j}=g\frac{\mathbf{v}^\top}{||\mathbf{v}||} {\rm{tanh}}(\mathbf{W}_{1}\mathbf{q}_{i\!-\!1}+\mathbf{W}_{2}\mathbf{h}_{j}+\mathbf{b})+r.\label{mce}
\end{equation}
The attention network learns the appropriate scale and offset of $e_{i,j}$ via trainable scalars $g$ and $r$. We have to initialize $r$ by setting it to a negative value to guarantee that the mean of $p_{i,j}$ approaches to 0 \cite{raffel2017online}. However, according to the term $p_{i,j}\mathbb{E}[z_{i\!-\!1,j}]$, the expectations of $z$ exponentially decay as $p_{i,j}$ along the index $i$, which indicates that the alignments disappear when predicting output labels at long distance. Therefore, $\mathbb{E}[z]$ tend to vanish along either $j$ or $i$. \par

In addition to the attention weights vanishing problem, HMA also suffers from the mismatch problem between the training and decoding stages. In the decoding stage, we replace the Bernoulli sampling with an indicator function $z_{i,j}=\mathbb{I}(p_{i.j}\!>\!0.5)$ assuming that $p_{i,j}$ is discrete, i.e., $p_{i,j}$ approaches 0 or 1. However, there is no strict guarantee that $p_{i,j}$ is close to 0 or 1. \par

\subsection{Monotonic Chunk-wise Attention (MoChA)}
MoChA extends HMA by aligning each output label to consecutive input representations, as shown in figure~\ref{fig:att} (b). The choice of chunk size is task-specific. \par

\subsubsection{Decoding of MoChA}
In the decoding stage, MoChA searches for end-points in the same way as HMA does. Additionally, MoChA preforms soft alignments on consecutive input representation vectors within a fixed-length chunk. Assuming that the end-point at the $(i-1)$-th output time-step is the $t_{i\!-\!1}$-th input representation vector $\mathbf{h}_{t_{i\!-\!1}}$, the next end-point $t_{i}$ is determined by (\ref{mhae})-(\ref{mhaz}), and the label-wise representation vector $\mathbf{r}_{i}$ is computed as follows, for $k=t_{i}\!-\!w\!+\!1,\cdot\cdot\cdot,t_{i}$:
\begin{eqnarray}
     u_{i,k}&=&\mathbf{v}^\top{\rm{tanh}}(\mathbf{W}_{1}\mathbf{q}_{i-1}+\mathbf{W}_{2}\mathbf{h}_k+\mathbf{b}),\label{mcu}\\
    \alpha_{i,k}&=&\left.{\rm{exp}}(u_{i,k})\middle/\sum_{l=t_{i}\!-\!w\!+\!1}^{t_{i}}{\rm{exp}}(u_{i,l})\right.,\label{mca}\\
    \mathbf{r}_{i}&=&\sum_{k=t_{i}\!-\!w\!+\!1}^{t_{i}}\alpha_{i,k}\mathbf{h}_{k},\label{mcc}
\end{eqnarray}
where matrices $\mathbf{W}_{1}$, $\mathbf{W}_{2}$ and bias vectors $\mathbf{b}$, $\mathbf{v}$ are learnable parameters. In (\ref{mcu}), $u_{i,k}$ is the pre-softmax activations. In (\ref{mca}), $w$ denotes the chunk width and $\alpha_{i,k}$ denotes the attention weight within the chunk. \par

The decoding computational complexity of MoChA is $\mathcal{O}(T\!+\!wL)$. Therefore, MoChA needs a less decoding computational cost when $T$, $L$ are large and $w$ is small, compared with the global attention models.
\subsubsection{Training of MoChA}
Similar to the training of HMA, we have to compute the expectation values of $\alpha_{i,j}$ and $\mathbf{r}_{i}$ when we train MoChA:
\begin{eqnarray}
    \mathbb{E}[\alpha_{i,j}]&=&\sum^{j\!+\!w\!-\!1}_{k=j}\frac{\mathbb{E}[z_{i,k}]\cdot{\rm exp}(u_{i,j})}{\sum^k_{l=k\!-\!w\!+\!1}{\rm exp}(u_{i,l})},\label{mcr}\\
    \mathbb{E}[\mathbf{r}_{i}]&=&\sum_{j=1}^{T}\mathbb{E}[\alpha_{i,j}]\cdot \mathbf{h}_{j},\label{mctc}    
\end{eqnarray}
where $\mathbb{E}[z_{i,k}]$ is the expectation value computed by (\ref{mha}). In (\ref{mctc}), $\mathbb{E}[\mathbf{r}_{i}]$ is the weighted sum of all the input representation vectors. \par

The vanishing problem of $\mathbb{E}[z]$ is still a tricky issue for MoChA and HMA. Similar to the findings in \cite{online-trans, google-long-utt}, we failed to train the MoChA and HMA with long utterances, shown in figure~\ref{fig:att_ws}(a)-(d), but succeeded with short utterances, shown in figure~\ref{fig:att_ws}(g)-(i).
It should be noted that the MoChA also suffers from the mismatch problem between the training and decoding stages. The work in \cite{he2019robust} found that this mismatch problem would harm the performance of MoChA. \par

\section{Proposed Online CTC/attention Based End-to-End ASR Architecture}
In this section, we propose several innovative algorithms to develop an online joint CTC/attention based end-to-end ASR architecture, including sMoChA, MTA, T-CTC prefix score and DWJD algorithms. \par

\subsection{Stable MoChA (sMoChA)}
We design sMoChA to address the vanishing attention weights problem of HMA and MoChA in the first part. Then, we alleviate the training-and-decoding mismatch problem of HMA and MoChA in the second part. \par

\subsubsection{Addressing Vanishing Attention Weights Problem}
Training sMoChA is different from training MoChA because we allow sMoChA to inspect the sequential input representation from the first element, rather than from the previous end-point like MoChA does. Let $i$ and $j$ denote the indices of output labels and input representation vectors, respectively. The expectation of the discrete variable $z_{i,j}$ changes to:
\begin{eqnarray}
    \mathbb{E}[z_{i,j}]&=&p_{i,j}\prod_{k=1}^{j-1}(1-p_{i,k})\\
    &=&\frac{p_{i,j}}{p_{i,j\!-\!1}}\cdot(1-p_{i,j\!-\!1})\mathbb{E}[z_{i,j\!-\!1}].\label{cz}    
\end{eqnarray}
\par
Now, we discuss the convergence property of sMoChA. In (\ref{cz}), $\mathbb{E}[z_{i,j}]$ no longer decays to zero with the increase of $i$, because $\mathbb{E}[z_{i,j}]$ is independent of the $\mathbb{E}[z_{i\!-\!1,j}]$. However, $\mathbb{E}[z_{i,j}]$ still exponentially decays to zero with the increase of $j$. To prevent $\mathbb{E}[z_{i,j}]$ from vanishing, we enforce the mean of $(1-p_{i,j})$ to be close to 1 by initializing $r$ in (\ref{mce}) to a negative value, e.g., $r=-4$. As illustrated in  figure~\ref{fig:att_ws} (e), sMoChA has successfully addressed the vanishing attention weights problem of the original MoChA. \par

\subsubsection{Alleviating Mismatch Problem in Training / Decoding}
To enable sMoChA applicable to online tasks, we force the end-point to move forward and perform alignments within a single chunk in the decoding stage, just like MoChA. However, there exists a mismatch problem between the training and decoding stages, because the label-wise representation vector $\mathbf{r}_{i}$ in the decoding stage is computed within a single chunk, while the $\mathbf{r}_{i}$ in the training stage is computed over all the input representation vectors. 
This mismatch problem will cause speech recognition accuracy degradation. To alleviate this mismatch problem, we propose to use the \textit{higher order decoding chunks mode} for sMoChA, in the decoding stage, rather than one single decoding chunk. When switching to the higher order decoding chunks mode, the formulation of $\mathbf{r}_{i}$ changes to:
\begin{eqnarray}
    \mathbb{\overline{E}}[z_{i,k}]&=&\left.\mathbb{E}[z_{i,k}]\middle/\sum_{l=t_{i}\!-\!n\!+\!1}^{t_{i}}\mathbb{E}[z_{i,l}]\right.,\label{smochanorm}\\
    \alpha_{i,j}&=&\sum^{j\!+\!w\!-\!1}_{k=j}\frac{\mathbb{\overline{E}}[z_{i,k}]{\rm exp}(u_{i,j})}{\sum^k_{l=k\!-\!w\!+\!1}{\rm exp}(u_{i,l})},\\
    \mathbf{r}_{i}&=&\sum_{j=t_{i}\!-\!w\!+\!n}^{t_{i}}\alpha_{i,j} \mathbf{h}_{j},
\end{eqnarray}
where $t_{i}$ denotes the selected end-point, $n$ denotes the chunk order, i.e., the number of consecutive decoding chunks, and $w$ denotes the chunk width. $\mathbb{\overline{E}}[z_{i,k}]$ is the normalization of $\mathbb{E}[z_{i,k}]$ across the consecutive chunks. $u_{i,j}$ is the pre-softmax activations computed by (\ref{mcu}), $\alpha_{i,j}$ is the attention weight and $\mathbf{h}_{j}$ is the $j$-th input representation vector.
Because of the higher order decoding chunks mode, sMoChA actually attends to $w\!+\!n\!-\!1$ consecutive input representation vectors at each time-step. The gap between the training and decoding stages is reduced with the increase of $n$. \par

The decoding computational complexity of sMoChA is as follows: (1) $\mathcal{O}(T\!+\!wL)$ for $n=1$, which is the same with MoChA; (2) $\mathcal{O}(TL\!+\!wL\!+\!nL)$ for $n>1$ because $\{p_{i,j\leq k}\}$ are required to compute $\mathbb{E}[z_{i,k}]$.
For $n\to\infty$, as shown in figure~\ref{fig:att}(c), sMochA covers all the historical input representation vectors. As $n$ increases, the decoding computational complexity increases. \par

\subsection{Monotonic Truncated Attention (MTA)}
During decoding, sMoChA suffers from the training-and-inference mismatch problem, and the receptive field of sMoChA is restricted to the predefined chunk width. 
Even though we can adopt the higher order decoding chunks mode for sMoChA, this bears the higher computational cost.
To better exploit the historical information and simplify sMoChA, we propose MTA to perform attention over the truncated historical input representations, as shown in figure~\ref{fig:att} (d). \par

\subsubsection{Decoding of MTA}
In the decoding stage, MTA starts the search from the previous end-point and selects a new end-point to truncate the sequential input representation. Then, MTA performs alignments on the truncated sequential input representation. Since MTA only needs truncated historical information, it can be applied to online tasks. Assuming that the maximum number of the input representation vectors is $T$, and the previous end-point at the ($i\!-\!1$)-th output time-step is $t_{i\!-\!1}$, we compute the attention weights for $j=1,\cdot\cdot\cdot, T$ as follows:
\begin{eqnarray}
    e_{i,j}&=&{\rm{Energy}}(\mathbf{q}_{i-1},\mathbf{h}_{j}),\\
    p_{i,j}&=&{\rm{Sigmoid}}(e_{i,j}),\label{sp}\\
    z_{i,j}&=&\mathbb{I}(p_{i,j}>0.5\ \wedge\ j\geq t_{i\!-\!1}),\label{ind}\\
    \alpha_{i,j}&=&p_{i,j}\prod_{k=1}^{j-1}(1-p_{i,k}),\label{mtaw}
\end{eqnarray}
where $\mathbf{q}_{i-1}$ is the $(i-1)$-th state of the decoder, $\mathbf{h}_{j}$ is the $j$-th input representation vector. The \emph{Energy} function will assign a larger value when $\mathbf{q}_{i-1}$ and $\mathbf{h}_{j}$ are more relative. $p_{i,j}$ is the truncation probability, $z_{i,j}$ is the discrete truncate or do not truncate decision and $\alpha_{i,j}$ is the attention weight. Once $p_{i,j}\!>\!0.5$ and $j\!\geq\!t_{i\!-\!1}$, MTA will stop computing attention weights and set the current end-point $t_{i}$ to $j$. MTA ensures that the end-point moves forward consistently by enforcing $t_{i}\!\geq\!t_{i\!-\!1}$. Then, the label-wise representation vector $\mathbf{r}_{i}$ is computed as follows: 
\begin{equation}
    \mathbf{r}_{i}=\sum_{j=1}^{t_{i}}\alpha_{i,j} \mathbf{h}_{j}.\label{mtai}
\end{equation}
\par
Equation (\ref{mtai}) shows that MTA includes all the historical input representation vectors, which means that MTA has broader receptive field and better modeling capability than HMA, MoChA and sMoChA with limited chunk order. MTA also simplifies the method of computing attention weights by using truncation probabilities straightly while MoChA and sMoChA use \emph{ChunkEnergy} function to recompute attention weights. \par

The decoding computational complexity of MTA is $\mathcal{O}(TL)$, where $T$ and $L$ denotes the lengths of input representation vectors and output labels, respectively. \par

\subsubsection{Training of MTA}
In the training stage, MTA discards the indicator function in (\ref{ind}) and computes the label-wise representation vector $\mathbf{r}_{i}$ as follows:
\begin{equation}
    \mathbf{r}_{i}=\sum_{j=1}^{T}\alpha_{i,j} \mathbf{h}_{j}.\label{mtat}
\end{equation}
Similarly, we choose (\ref{mce}) as the \emph{Energy} function and initialize the bias $r$ to a negative value, e.g., $r=-4$, just like sMoChA, to prevent the attention weights from vanishing. Furthermore, MTA alleviates the training-and-decoding mismatch problem. According to (\ref{mtaw}) and (\ref{mtat}), MTA performs alignments on top of all the input representation vectors in the training stage, but the attention weight $\alpha_{i,j}$ after the end-point is close to 0 due to the sharp selection probability $p_{i,j}$ at the end-point. Therefore, there is little influence on the recognition accuracy when the future information is unavailable in the decoding stage. \par

\subsection{Truncated CTC (T-CTC) Prefix Score}
The joint CTC/attention decoding is of crucial importance for the decoder to generate high quality hypotheses. However, as we have stressed in section \uppercase\expandafter{\romannumeral2}-B, the CTC prefix scores are computed on the full utterance. In this section, we propose a truncated CTC (T-CTC) prefix score to stream the calculation of CTC prefix score. \par

\begin{algorithm}[t]
    \footnotesize
    \caption{\footnotesize Truncated CTC Prefix Score Algorithm}
    \label{t-ctc}
    \KwIn{output label index $i$, input representation index $j$, total number of input representation vectors $T_{\rm max}$, $n$-length partial hypothesis $l=(y_{1}\!=\!\langle sos\rangle, y_{2}, \cdot\cdot\cdot, y_{n})$, CTC probability distribution $\{p(y_{i}|\mathbf{h}_{j})\}$ with respect to input representation vector $\mathbf{h}_{j}$, previous end-point $t_{n\!-\!1}$, threshold $\theta$.}
    \KwOut{truncated CTC prefix score $S_{\rm{tctc}}$ for $l$, current end-point $t_{n}$.}
    \eIf{$y_{n}=\langle eos\rangle$}{
        $S_{\rm{tctc}}={\rm{log}}\{\gamma^{n}_{t_{n\!-\!1}}(l_{1:n\!-\!1})+\gamma^{b}_{t_{n\!-\!1}}(l_{1:n\!-\!1})\}$\;
        $t_{n}=t_{n\!-\!1}$\;
    }
    {
        $\gamma^{n}_{1}(l)=\left\{
        \begin{array}{ccl}
        p(y_{n}|\mathbf{h}_{1}) & & {\rm{if}}\; l_{1:n\!-\!1}=\langle sos\rangle\\
        0 & & {\rm{otherwise}}
        \end{array} \right.$\;
        $\gamma^{b}_{1}(l)=0$\;
        $\Psi=\gamma^{n}_{1}(l)$\;
        $j=2$\;
        \While{$j\leq T_{\rm max}$}
        {
            \If{$j>t_{n\!-\!1}$}
            {
                $\gamma^{n}_{j}(\langle sos\rangle)=0$\;
                $\gamma^{b}_{j}(\langle sos\rangle)=\gamma^{b}_{j\!-\!1}(\langle sos\rangle)\cdot p(\langle b\rangle|\mathbf{h}_{j})$\; 
                \For{$i=2,\cdot\cdot\cdot,n\!-\!1$}{
                    $\Phi=\gamma^{b}_{j\!-\!1}(l_{1:i\!-\!1})+\left\{
                    \begin{array}{cl}
                    0 & {\rm{if}}\; y_{i\!-\!1}=y_{i}\\
                    \gamma^{n}_{j\!-\!1}(l_{1:i\!-\!1}) & {\rm{otherwise}}
                    \end{array} \right.$\;
                    $\gamma^{n}_{j}(l_{1:i})=\left(\gamma^{n}_{j\!-\!1}(l_{1:i})+\Phi\right)\cdot p(y_{i}|\mathbf{h}_{j})$\;
                    $\gamma^{b}_{j}(l_{1:i})=\left(\gamma^{b}_{j\!-\!1}(l_{1:i})+\gamma^{n}_{j\!-\!1}(l_{1:i})\right)\cdot p(\langle b\rangle|\mathbf{h}_{j})$\;        
                }                
            }
            $\Phi=\gamma^{b}_{j\!-\!1}(l_{1:n\!-\!1})+\left\{
            \begin{array}{ccl}
            0 & & {\rm{if}}\; y_{n\!-\!1}=y_{n}\\
            \gamma^{n}_{j\!-\!1}(l_{1:n\!-\!1}) & & {\rm{otherwise}}
            \end{array} \right.$\;
            $\gamma^{n}_{j}(l)=\left(\gamma^{n}_{j\!-\!1}(l)+\Phi\right)\cdot p(y_{n}|\mathbf{h}_{j})$\;
            $\gamma^{b}_{j}(l)=\left(\gamma^{b}_{j\!-\!1}(l)+\gamma^{n}_{j\!-\!1}(l)\right)\cdot p(\langle b\rangle|\mathbf{h}_{j})$\;
            $\Psi=\Psi+\Phi\cdot p(y_{n}|\mathbf{h}_{j})$\;
            \eIf{$j>t_{n\!-\!1}$ {\rm and} $\Phi\cdot p(y_{n}|\mathbf{h}_{j})<\theta$}
            {
                \textbf{break}\;
            }
            {
                $j=j+1$\;            
            }
        }
        $t_{n}=j$\;
        $S_{\rm{tctc}}={\rm{log}}\Psi$\;
    }
\end{algorithm}

\subsubsection{Properties of CTC Prefix Scores}
First, we discuss the reason why it is redundant to compute CTC prefix score based on the full utterance. Given that the encoder has generated a sequential input representation $H=(\mathbf{h}_1, \cdot\cdot\cdot, \mathbf{h}_{T_{\rm max}})$ of length $T_{\rm max}$ and the decoder has generated a partial hypothesis $l\!=\!(y_{1}, \cdot\cdot\cdot,y_{n})$ of length $n$, the probability of $l$ over $H_{1:j}$ in (\ref{ctcsum}) is computed by:
\begin{equation}
    p_{\rm{ctc}}(l|H_{1:j})=\Phi\cdot p(y_{n}|\mathbf{h}_{j}),
\end{equation}
\begin{equation}
    \Phi=\left\{
    \begin{array}{ll}
     \gamma^{b}_{j\!-\!1}(l_{1:n\!-\!1}) &  {\rm{if}}\; y_{n\!-\!1}=y_{n}\\
     \gamma^{b}_{j\!-\!1}(l_{1:n\!-\!1})+\gamma^{n}_{j\!-\!1}(l_{1:n\!-\!1}) & {\rm{otherwise}} 
    \end{array}
    \right. ,
\end{equation}
where $p(y_{n}|\mathbf{h}_{j})$ is the $j$-th output of the CTC branch, which also represents the posterior probability. $\gamma^{n}_{j}(l)$ and $\gamma^{b}_{j}(l)$ denote the forward probabilities of $l$ over $H_{1:j}$, where the superscripts $n$ and $b$ represent different cases in which all CTC alignments end with a non-blank or blank label. \par

\begin{figure}[t]
    \centering
    \includegraphics[scale=0.5]{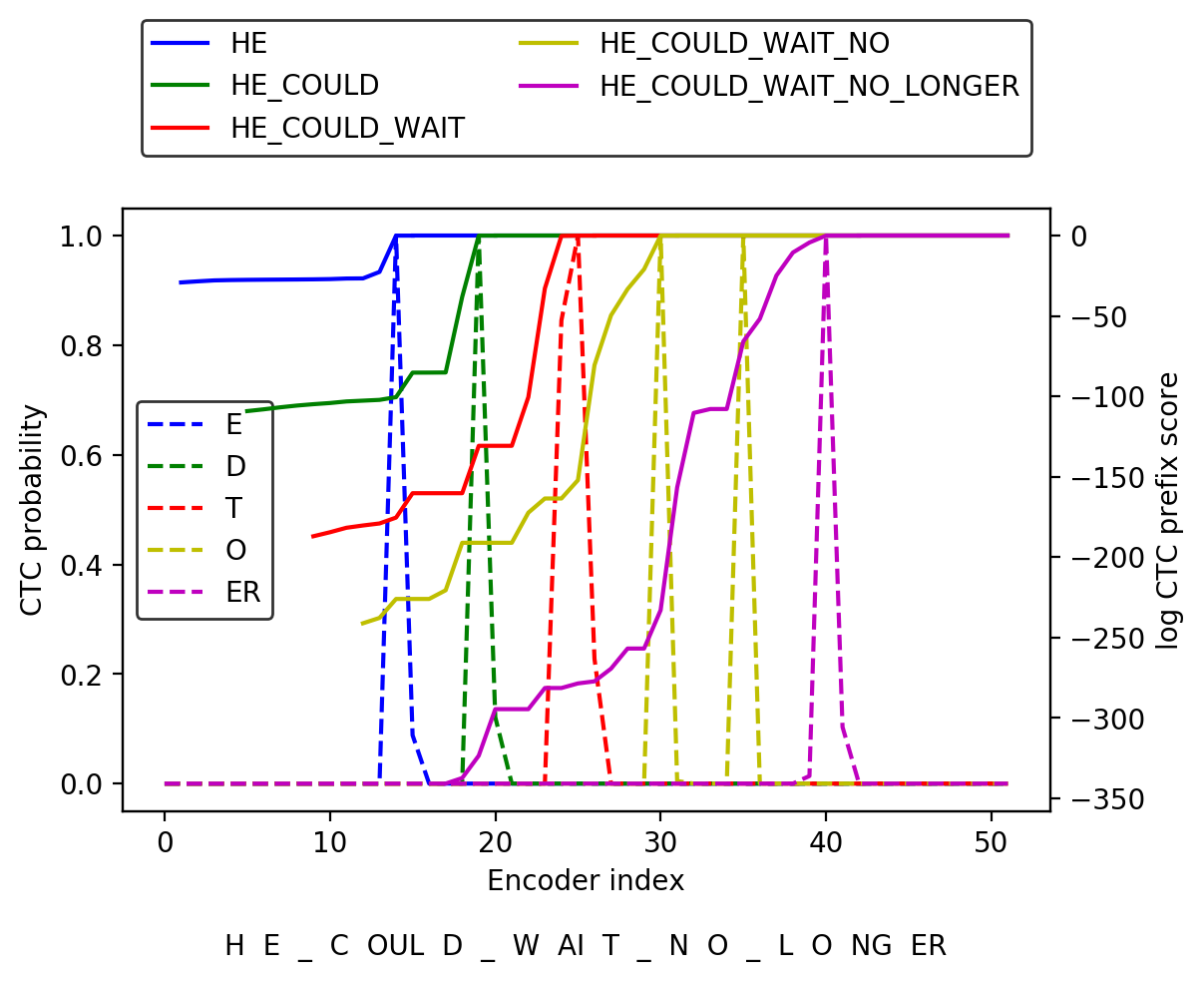}
    \caption{Illustration of CTC prefix scores. We use the pronunciation-assisted sub-word modeling (PASM) \cite{xu2019improving} method to generate CTC/attention model output labels. For simplicity, we only show partial hypotheses at word endings. The solid lines represent CTC prefix scores and the dashed lines represent CTC posterior probabilities of sub-words at word endings.}
    \label{fig:tctc}
\end{figure}

Considering the peaky posterior properties of CTC-based models \cite{Chen2017Phone}, $p(y_{n}|\mathbf{h}_{j})$ is approximately equal to 0 everywhere except for the time-steps when the CTC-based model predicts $y_{n}$. Therefore, only when the CTC-based model predicts $l$ over $H_{1:j}$ and predicts $y_{n}$ at $\mathbf{h}_{j}$ for the first time, $p_{\rm{ctc}}(l|H_{1:j})\gg0$; otherwise  $p_{\rm{ctc}}(l|H_{1:j})\approx0$. We can always find an end-point $t_{n}$ that $p_{\rm{ctc}}(l|H_{1:j})\approx0$ for $j\!>\!t_{n}$. Figure~\ref{fig:tctc} illustrates the property of CTC prefix scores, which is in line with our analysis. Based on this property, we only need $t_{n}$ input representation vectors to estimate the CTC prefix score of $l$. The proposed T-CTC prefix score is thus computed as follows:
\begin{equation}
    S_{\rm{tctc}}={\rm{log}}\sum_{j=1}^{t_{n}}p_{\rm{ctc}}(l|H_{1:j}).\label{eq-tctc}
\end{equation}
\par
\subsubsection{T-CTC Prefix Score Algorithm}
The Algorithm~\ref{t-ctc} describes how the CTC branch determines the end-point $t_{n}$ and computes the T-CTC prefix score of $l$ over $H_{1:t_{n}}$. 
The initialization and recursion steps for the forward probabilities, $\gamma^{n}_{j}(l)$ and $\gamma^{b}_{j}(l)$, are described in lines 5-7 and lines 9-28, respectively. When $j\leq t_{n\!-\!1}$, the T-CTC prefix probability $\Psi$ can be updated directly (line 22) because $\gamma^{n}_{j\!-\!1}(l_{1:n\!-\!1})$ and $\gamma^{b}_{j\!-\!1}(l_{1:n\!-\!1})$ have already been calculated in the previous step. When $j>t_{n\!-\!1}$, the missing forward probabilities have to be computed before updating $\Psi$ (lines 10-18). When the change of $\Psi$ is less than the threshold $\theta$, e.g., $\theta\!=\!10^{-8}$, the recursion step will be terminated (line 24). The algorithm enforces the end-point to move forward as the hypothesis expands, which is consistent with monotonic alignments in CTC-based models.
Finally, the algorithm returns a T-CTC prefix score when the last label of $l$ is $\langle eos\rangle$ (lines 1-3). It should be noted that the end-point will not exceed the total number of input representation vectors $T_{\rm max}$, which is determined by voice activity detection (VAD) as the input frames comes. \par

\subsubsection{T-CTC vs. CTC}
The T-CTC prefix scores not only approximates the CTC prefix score, but also have lower computational complexity than the CTC prefix scores. For a partial hypothesis with the end-point $t_{n}$, the T-CTC prefix score algorithm reduces the computational cost to $t_{n}/T_{\rm max}$. During the beam search, the majority of partial hypotheses are pruned and the T-CTC prefix score algorithm can accelerate the joint CTC/attention decoding process. \par

\subsection{Dynamic Waiting Joint Decoding (DWJD)}
\begin{figure}[t]
    \centering
    \includegraphics[width=\linewidth]{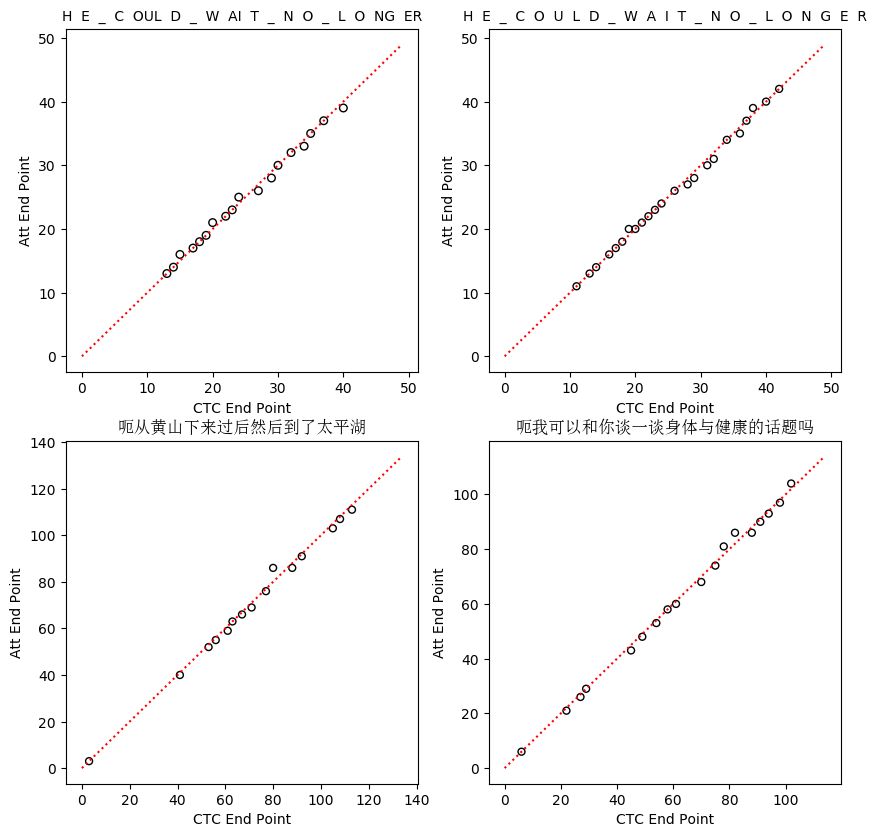}
    \caption{Illustration of unsynchronized predictions of CTC and MTA. Each black circle denotes a partial hypothesis. The horizontal and vertical coordinate represent the end-points determined by CTC and MTA, respectively. The red dotted line represent synchronous predictions. Some black circles deviate from the red dotted line, which means that CTC and MTA have the disagreement on where to truncate the input representations.}
    \label{fig:dwjd}
\end{figure}

We have described the proposed streaming attention methods and T-CTC prefix score algorithm in previous sections. However, it remains a problem that how the encoder, the attention-based decoder and the CTC branch collaborate to generate and prune the hypotheses in the online decoding stage. Because the attention based decoder and the CTC branch do not generate synchronized predictions, as shown in figure~\ref{fig:dwjd}, we have to compute the scores of the same hypothesis in the attention-based decoder and the CTC branch at different time-steps. To address the CTC/attention unsynchronized prediction problem in the online setting, we propose a dynamic waiting joint decoding (DWJD) algorithm, which dynamically collects the prediction scores of the attention-based decoder and the CTC branch in an online manner. See Algorithm~\ref{DWJD} for the details of DWJD algorithm. \par

\begin{algorithm}[t]
    \footnotesize
    \caption{\footnotesize Dynamic Waiting Joint Decoding (DWJD)}
    \label{DWJD}    
    \KwIn{input frames $X=\{x_{1}, x_{2}, \cdot\cdot\cdot\}$, output index $i$, input representation index $j$, total number of input representation vectors $T_{\rm max}$, input representation vector $\mathbf{h}_{j}$, previous state of the decoder $\mathbf{q}_{i\!-\!1}$, label-wise representation vector $\mathbf{r}_{i}$, output label $y_{i}$, end-points $t_{\rm att}$, $t_{\rm ctc}$ and $t_{\rm enc}$, sigmoid function $\sigma(\cdot)$, threshold $\theta$.}
    $\mathbf{q}_{0}\!=\!\vec{0},\ y_{0}\!=\!l=\!\langle sos\rangle,\ t_{\rm att}\!=\!t_{\rm ctc}\!=\!t_{\rm enc}\!=\!1,\ i\!=\!1,\ \theta\!=\!10^{-8}$\;
    $\mathbf{h}_{1}={\rm Encoder}(X)$\;
    \While{$y_{i-1}\ne\langle eos\rangle$}{
        $j=t_{\rm att}$\;
        \While{$j\leq T_{\rm max}$}{
            \If{$j>t_{\rm enc}$}{
                $\mathbf{h}_{j}={\rm Encoder}(X)$, $t_{\rm enc}=t_{\rm enc}+1$\;
            }
            $p_{i,j}=\sigma({\rm Energy}(\mathbf{q}_{i-1}, \mathbf{h}_{j}))$\;
            \eIf{$p_{i,j}\geq0.5$}{
                $t_{\rm att}=j$\;
                $\mathbf{r}_{i}={\rm sMoChA}(\mathbf{q}_{i-1}, H_{1:t_{\rm att}})$\;
                \textbf{break}\;
            }
            {
                $j=j+1$\;
            }
        }
        \If{$j> T_{\rm max}$}{
            $\mathbf{r}_{i}=0$\;
        }
        $y_{i}\sim{\rm Decoder}(\mathbf{r}_{i}, \mathbf{q}_{i-1}, y_{i-1}$)\;
        $l=l\cdot y_{i},\quad i=i+1,\quad j=1$\;
        \While{$j\leq T_{\rm max}$}{
            \If{$j>t_{\rm enc}$}{
                $\mathbf{h}_{j}={\rm Encoder}(X)$, $t_{\rm enc}=t_{\rm enc}+1$\;
            }
            Get $p_{\rm{ctc}}(l|H_{1:j})$\;
            \eIf{$j>t_{\rm ctc}$ {\rm and} $p_{\rm{ctc}}(l|H_{1:j})<\theta$}{
                $t_{\rm ctc}=j$\;
                \textbf{break}\;
            }
            {
                $j=j+1$\;
            }
        }
        Compute $S_{\rm tctc},\ S_{\rm att}$\;  
    }
\end{algorithm}

\subsubsection{DWJD Algorithm}
In the joint CTC/attention decoding method, the attention based decoder generates new hypotheses and the CTC branch assists in pruning the hypotheses. In the DWJD algorithm, the streaming attention model first searches the end-point $t_{\rm att}$ and compute the label-wise representation vector $\mathbf{r}_{i}$ based on $t_{\rm att}$ input representation vectors (lines 4-20). After that, the decoder predicts a new label, along with the corresponding decoder score $S_{\rm att}$ (lines 21-22). Then, the CTC branch searches the end-point $t_{\rm ctc}$ and computes the T-CTC prefix score $S_{\rm tctc}$ of the new hypothesis conditioned on $t_{\rm ctc}$ input representation vectors (lines 23-34). Finally, we prune this hypothesis by the combination of $S_{\rm att}$ and $S_{\rm tctc}$. In our experiments, we also utilize a separately trained LSTM language model, together with $S_{\rm ctc}$ and $S_{\rm att}$, to prune $l$ in the beam search. \par

In the offline joint CTC/attention decoding method, the forward propagation of the encoder and the beam search of the decoder perform separately. However, in our online joint CTC/attention decoding method, the encoder and decoder perform concurrently and have to be carefully scheduled. 
The key of the DWJD algorithm is to organize the input representation vectors, i.e., streaming encoder outputs, in the beam search. When the attention-based decoder and the CTC branch predict the end-points at different time-steps, or different hypotheses correspond to different end-points, the hypothesis with the  end-point on the back should be suspended until the encoder provides enough outputs. For example, if the attention-based decoder needs more encoder outputs than the CTC branch, the attention-based decoder has to wait for the encoder to produce enough outputs in the first place. When the decoding process switch to the CTC branch, we just need to retrieve the encoder outputs obtained. \par 

\subsubsection{End Detection Criteria}
In our online joint CTC/attention decoding method, it is important to terminate the beam search appropriately. We find that the T-CTC prefix score cannot exclude the premature ending hypotheses during the beam search as the CTC prefix score does. If the decoder predicts $\langle eos\rangle$ earlier than expected, T-CTC will assign a higher score than the CTC prefix score because such hypotheses are reasonably based on the truncated information, but may be wrong conditioned on the complete utterances. Besides, suppose the hypotheses have the same prefix, the decoder and the external language model assign higher scores to shorter hypotheses because of the property of probabilities ($0\!\leq\!p\!\leq\!1$). Therefore, the scores of short hypotheses tend to be higher than that of long hypotheses in our online joint CTC/attention decoding method. The end detection criteria adopted in \cite{watanabe2017hybrid}, which terminates the beam search when the scores get lower, will lead to the early stop problem in our online system. To tackle this problem, we design a different end detection criteria to terminate the decoder appropriately. First, we believe that the monotonic attention based decoder should continue predicting new labels until the end-point $t_{\rm ctc}$ reaches the last encoder outputs, which is determined by VAD in the front-end. Therefore, our online joint CTC/attention decoding method can get rid of the requirements for length penalty factor, coverage term \cite{Chorowski2017Towards} and other predefined terms that are often adopted in label-synchronous decoding. Second, we use the end detection criteria as follows:
\begin{equation}
    \sum^{M}_{m=1}\mathbb{I}\left(\max_{l\in\Omega:|l|=n}S(l)-\max_{l^{'}\in\Omega:|l^{'}|=n\!\!-m}S(l^{'})<D_{\rm end}\right)=M,\label{newcrit}
\end{equation}
given that the length of current complete hypotheses is $n$. Equation (\ref{newcrit}) becomes true if scores of current complete hypotheses are much smaller than those of shorter complete hypotheses, which implies that there is little chance to get longer hypotheses with higher scores. In our experiments, we set $M$ to 3 and $D_{\rm end}$ to -10. For the final recognized hypotheses, we suggest to replace the T-CTC prefix scores of the collected ending hypotheses with the CTC prefix score to eliminate those premature ending hypotheses. \par

\section{Experiments and Results}
\subsection{Corpus}
We evaluated our experiments using two different ASR tasks. The first ASR task uses the standard English speech corpus, LibriSpeech \cite{panayotov2015librispeech} (about 960h). The second ASR task is HKUST Mandarin conversational telephone (HKUST) \cite{liu2006hkust} (about 200h). Experiments with these two corpora were designed to show the effectiveness of the proposed online methods. To improve the ASR accuracy in HKUST task, we applied speed perturbation \cite{Ko2015_speed_perturb}, of which the speed perturbation factors were 0.9, 1.0 and 1.1, to the training sets of HKUST. The development sets of both LibriSpeech and HKUST were used to tune the ASR model training and decoding hyperparameters. \par

\renewcommand{\arraystretch}{1.5}
\begin{table}
    \centering
    \caption{Experimental hyperparameters for CTC/attention models. }
    \scriptsize
    \begin{tabularx}{0.95\linewidth}{p{0.525\linewidth}| c| c}
    \Xhline{2\arrayrulewidth}
    Parameter & LibriSpeech & HKUST \\
    \hline
	 CTC Weight for Training ($\lambda$)&0.5 &0.5\\
    \hline
    Batch Size & 20 & 20 \\    
    \hline
    Epoch & 10 & 15 \\
    \hline
    \multirow{2}{*}{Adadelta\cite{zeiler2012adadelta} Optimizer} &
    $\rho\!=\!0.95$ & 
    $\rho\!=\!0.95$ \\
    &
    $\epsilon\!=\!10^{-8}$ &
    $\epsilon\!=\!10^{-8}$ \\
    \hline
    Gradient Clipping Norm & 5 & 5\\
    \hline
    CTC Weight for Decoding ($\mu$)&0.5 &0.6\\
    \hline   
    Language Model Weight ($\beta$)& 0.7 & 0.3 \\
    \hline
    Decoding Beam Size & 20 & 20\\    
    \Xhline{2\arrayrulewidth}
    \end{tabularx}
    \label{model_h}
\end{table}

\begin{figure*}
    \centering
    \includegraphics[width=0.85\linewidth]{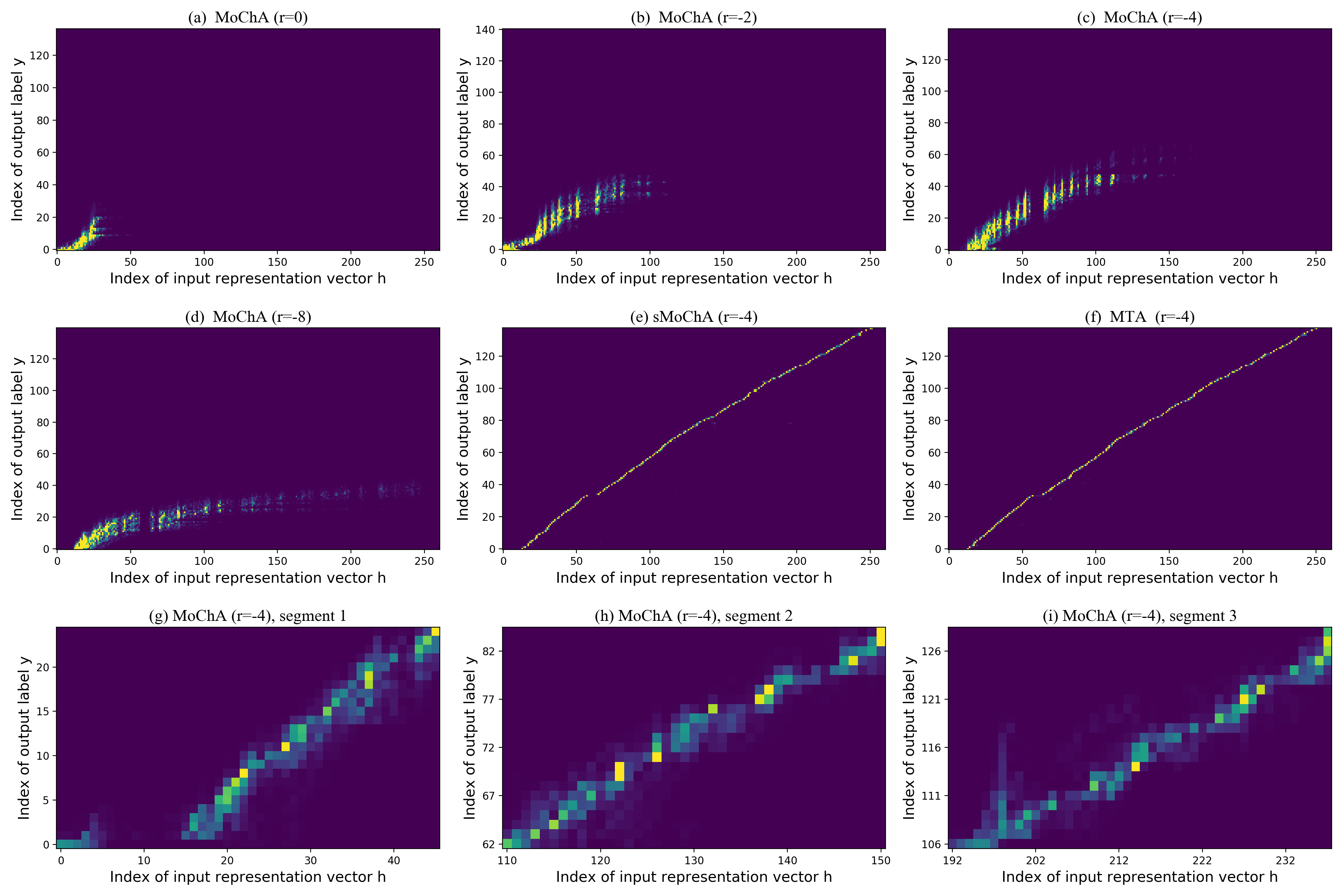}
    \centering\caption{Comparison of the training attention weights of the same utterance, which were learned by different streaming attention methods after 5 epochs training on LibriSpeech task. Figure(a)-(d) illustrate that the vanishing attention weights problem of MoChA found in long utterances. The tried bias initialization for MoChA included $r=0, -2, -4, -8$. Figure(e)-(f) display the training attention weights of sMoChA and MTA. From the comparisons, we can see that sMoChA/MTA addressed the vanishing attention weights problem and learned nearly monotonic alignments. We also choose three segments from the same utterance and show the training attention weights of the converged MoChA in figure(g)-(i).}
    \label{fig:att_ws}
\end{figure*}

\subsection{ASR Model Descriptions}
We used ESPNet \cite{Watanabe2018ESPnet} toolkit to build both the offline CTC/attention end-to-end ASR baselines and the proposed online models. \par

\subsubsection{Inputs}
For all ASR models, we used 83-dimensional features, which included 80-dimensional filter banks, pitch, delta-pitch and Normalized Cross-Correlation Functions (NCCFs). The features were computed with a 25 ms window and shifted every 10 ms. \par

\subsubsection{Outputs}
For LibriSpeech task, we tried English characters and pronunciation-assisted sub-word modeling (PASM) \cite{xu2019improving} units as the ASR model outputs. PASM is a sub-word extraction method that leverages the pronunciation information of a word. The characters consisted of 26 characters, and PASM units contained 200 sub-word units. In addition, both the character based and PASM based CTC/attention ASR models used four other output units, which included apostrophe, space, blank and sos/eos tokens. For HKUST task, we adopted a 3655-sized output set as the output units of the HKUST ASR model. The output units include 3623 Chinese characters, 26 English characters, as well as 6 non-language symbols denoting laughter, noise, vocalized noise, blank, unknown-character and sos/eos. \par

\begin{table*}[t]
    \centering
    \caption{Word error rate (WER) of different attention methods on LibriSpeech. \protect\\ Encoder types are VGG-BLSTM-Large (VBL) and VGG-BLSTM-Small (VBS).}
    \scriptsize
    \begin{tabular}{ c| c| c| c| c| c| c| c}
    \Xhline{2\arrayrulewidth}
     &
     &
     &
     &
    \multicolumn{4}{c}{WER (\%) of test-clean / other} \\
    \cline{5-8}
    Attention Method &
    Chunk &
    Chunk &
    Streaming &
    \multicolumn{2}{c|}{VBL} &
    \multicolumn{2}{c}{VBS} \\
    \cline{5-8}
     &
    Width &
    Order &
    Attention &
    Character &
    PASM &
    Character &
    PASM    \\
    \hline
    Location-aware attention (LoAA) &
    -- &
    -- &
    No &
    4.0 / 12.8 &
    3.9 / 12.2 &
    4.6 / 14.4 &
    4.4 / 14.2 \\
    \hline
    Hard Monotonic Attention (HMA) &
    1 &
    -- &
    \multirow{6}{*}{Yes} &
    6.6 / 18.4 &
    5.8 / 16.0 &
    8.6 / 20.3 &
    8.8 / 21.0 \\
    \cline{1-3}\cline{5-8}
    Monotonic Chunk-wise Attention (MoChA) &
    3 &
    -- &
     &
    6.6 / 17.7 &
    5.5 / 15.8 &
    8.7 / 21.9 &
    8.5 / 20.8 \\
    \cline{1-3}\cline{5-8}
    \multirow{3}{*}{Stable MoChA (sMoChA)} &
     1 &
     1 &
      &
     4.3 / 13.3 &
     4.9 / 13.1 &
     6.5 / 16.8 &
     8.0 / 17.5 \\
      \cline{2-3}\cline{5-8}
     &
    \multirow{2}{*}{3} &
    1 &
     &
    4.5 / 13.3 &
    4.2 / 12.8 &
    8.8 / 19.8 &
    7.8 / 16.9 \\
    \cline{3-3}\cline{5-8}
     &
     &
     $\infty$ &
     &
     4.2 / 13.1 &
     4.0 / 12.6 &
     4.8 / 14.6 &
     4.7 / 14.4 \\
      \cline{1-3}\cline{5-8}
      Monotonic trunctaed attention (MTA) &
      -- &
      -- &
      &
      4.1 / 13.1 &
      4.0 / 12.2 &
      4.7 / 14.4 &
      4.8 / 14.2 \\
    \Xhline{2\arrayrulewidth}
    \end{tabular}
    \label{libri}
\end{table*}

\subsubsection{CTC/attention Encoder and Decoder}
We used VGGNet-style CNN-blocks as the CTC/attention encoder front-end. Each CNN-block contained two CNN layers and one max-pooling layer. The CNN layers used ReLU activation function. The first CNN-block had 64 filter kernels for each CNN layer, and the second CNN-block had 128 filter kernels for each CNN layer. All the kernel size is 3. The max-pooling layer had a $2\!\times\!2$ pooling window with a $2\!\times\!2$ stride.
Following the CNN-blocks, a multi-layer BLSTM was used for offline systems and a multi-layer Uni-LSTM or LC-BLSTM was used for online systems. 
For LibriSpeech task, we used 5-layer BLSTM with 1024 cells for VGG-BLSTM-Large (VBL) and VGG-LC-BLSTM-Large (LC-VBL), 3-layer BLSTM with 640 cells for VGG-BLSTM-Small (VBS) and VGG-LC-BLSTM-Small (LC-VBS), 10-layer Uni-LSTM with 1024 cells for VGG-Uni-LSTM-Large (VUL) and 6-layer Uni-LSTM with 640 cells for VGG-Uni-LSTM-Small (VUS).
For HKUST task, we used 3-layer BLSTM with 1024 cells for VBL and LC-VBL, 3-layer BLSTM with 640 cells for VBS and LC-VBS, 6-layer Uni-LSTM with 1024 cells for VUL and 6-layer Uni-LSTM with 640 cells for VUS. After that, we used an 1-layer and 2-layer fully-connected neural network for large and small models, respectively. The output frame-rate of all models was 25 Hz. The CTC/attention decoder was one 2-layer Uni-LSTM with 1024 cells and 640 cells for large and small models, respectively. The parameter sizes of (LC-)VBL, (LC-)VBS, VUL, VUS for LibriSpeech task are 151M, 46M, 111M and 35M, respectively. The parameter sizes of (LC-)VBL, (LC-)VBS, VUL, VUS for HKUST task are 112M, 53M, 89M and 42M, respectively. The training and decoding parameters are listed in table \ref{model_h}. \par

\subsubsection{External Language Model}
We followed the language model configurations in ESPnet \cite{Watanabe2018ESPnet}. We applied a 1-layer Uni-LSTM with 1024 cells for Librspeech task and 2-layer Uni-LSTM with 640 cells for HKUST task. We trained the RNN language models separately from the CTC/attention ASR models. For LibriSpeech task, the text corpus contained the LibriSpeech training transcripts and 14500 public domain books. For HKUST task, the text corpus only consisted of the training transcripts. It should be noted that, for LibriSpeech task, we implemented the multi-level language model decoding method \cite{multi-level-rnnlm-shinji} for both the character based and PASM based joint CTC/attention decoding. \par

\subsection{Results of Streaming Attention}
First, we compare various streaming attention methods by observing their attention weights during training. From figure~\ref{fig:att_ws}, we can see that, HMA or MoChA failed to learn monotonic alignments on LibriSpeech. We tried different initialization biases and trained for more than 5 epochs, but we did not observe convergence of HMA or MoChA. As mentioned in section \uppercase\expandafter{\romannumeral3}-A, MoChA could have handled input representation vectors at longer distance if the initialization of bias $r$ in (\ref{mce}) was smaller. From figure~\ref{fig:att_ws}(a)-(d), we can see that the MoChA tended to attend over broader encoder span, but with the sacrifice that the attention weights vanished faster as the output increased, which is in line with our analysis in section \uppercase\expandafter{\romannumeral3}-A and other related work \cite{online-trans,google-long-utt}. Because 80\% utterances have more than 100 outputs in LibriSpeech when using PASM units, we have to constrain the length of output sequence by segmenting the utterances in accordance with a DNN/HMM system. In our experiments, the output sequence lengths of all utterances are within 25 when training and testing HMA and MoChA on LibriSpeech. From figure~\ref{fig:att_ws}(g)-(i), we can see that the MoChA did learn monotonic alignments in segmented utterances. Different from LibriSpeech, 85\% utterances in HKUST have less than 20 characters and the longest output sequence length is 64. Therefore, we observed the convergence of the HMA and MoChA on HKUST. However, our proposed streaming attention methods, i.e., sMoChA and MTA, overcome the difficulty of modeling long utterances. Figure~\ref{fig:att_ws}(e)-(f) show that sMoChA and MTA were stable and learned nearly monotonic alignments with the suggested initial bias $r=-4$ in \cite{chiu2017monotonic}. \par

Next, we compare the performance of the mentioned attention methods on LibriSpeech and HKUST. \par

\renewcommand{\arraystretch}{1.5}
 \begin{table}
   \caption{Character error rate (CER) of different \protect\\ attention methods on HKUST.}
  \scriptsize
  \center
  \begin{tabular}{ c| c| c| c| c| c}
    \Xhline{2\arrayrulewidth}
    Attention &
    Chunk &
    Chunk &
    Streaming &
    \multicolumn{2}{c}{CER (\%) of dev / eval} \\
    \cline{5-6}
    Method &
    Width &
    Order &
    Attention &
    VBL &
    VBS \\
    \hline
    LoAA &
    -- &
    -- &
    No &
    28.7 / 27.6 &
    29.5 / 28.2 \\
    \hline
    HMA &
    1 &
    -- &
    \multirow{9}{*}{Yes} &
    32.1 / 30.0 &
    32.6 / 30.2 \\
    \cline{1-3}\cline{5-6}
    \multirow{2}{*}{MoChA} &
    3 &
    -- &
     &
    32.9 / 30.6 &
    33.1 / 31.3 \\
    \cline{2-3}\cline{5-6}
     &
    6 &
    -- &
     &
    33.0 / 30.3 &
    33.5 / 31.1 \\
    \cline{1-3}\cline{5-6}     
    \multirow{5}{*}{sMoChA}  &
    1 &
    1 &
     &
    28.8 / 27.8 &
    30.3 / 28.6 \\
    \cline{2-3}\cline{5-6}
     &
     \multirow{2}{*}{3} &
    1 &
      &
    28.6 / 27.8 &
    30.2 / 29.0 \\
    \cline{3-3}\cline{5-6}
      &
      &
    $\infty$   & 
      &
    28.5 / 27.7 &
    29.3 / 28.5 \\
    \cline{2-3}\cline{5-6}
      &
    \multirow{2}{*}{6} &
    1 &
      &
    28.9 / 27.7 &
    30.4 / 29.0 \\
    \cline{3-3}\cline{5-6}
       &
       &
    $\infty$    &
       &
    28.6 / 27.6 &
    29.6 / 28.3 \\
    \cline{1-3}\cline{5-6}
    MTA &
    -- &
    -- &
       &
    28.5 / 27.6 &
    29.4 / 28.1 \\ 
    \Xhline{2\arrayrulewidth}
    \end{tabular}
    \label{hk}
\end{table}

\subsubsection{LibriSpeech}
Table~\ref{libri} lists experimental results on LibriSpeech task. Word error rate (WER) was used to measure the ASR accuracy. First, we trained the offline baselines by using both the English character output units and PASM output units. In the first line, we can see that PASM based models outperformed the character based models, and VBL models outperformed VBS models. The results proved that pronunciation information contributed to the English ASR tasks, and larger models could learn better. \par

Second, we evaluated HMA, MoChA and the proposed sMoChA in lines 2-6. It shows that sMoChA outperformed HMA and MoChA. Furthermore, we explored how chunk width and chunk order would affect the performance of sMoChA. When chunk order was set to be 1, we can see that the character-level models performed better with smaller chunk width ($chunk\ width = 1$), while the PASM-based model performed better with larger chunk width ($chunk\ width = 3$). It implied that a broader context was helpful for PASM units, which were always longer than the single English character. 
As mentioned in section \uppercase\expandafter{\romannumeral4}-A, higher decoding chunk order can reduce the mismatch phenomenon between sMoChA training and decoding, and thus stabilize and improve the decoding accuracy of sMoChA. We can see that, by increasing the chunk order from 1 to $\infty$, the sMoChA-based model achieved consistent ASR accuracy improvement. The accuracy improvements of VBS models were larger than those of VBL models. This might because VBS models were much smaller and more vulnerable to the mismatch between the training and decoding stages. \par

Finally, we evaluated the proposed MTA method. 
MTA does not have the concept of chunk width or chunk order, its training and decoding are designed to be matched. Compared with the $chunk\ order = \infty$ decoding mode of sMoChA, MTA needs a less computational cost. 
From table~\ref{libri}, we can see that, no matter what model sizes or what output units we tried, MTA performed almost the best among the streaming attention methods, except for the VBS model with PASM units on the test-clean test set of LibriSpeech task. Compared with the LoAA based offline baselines, the MTA based models achieved comparable performance. \par

\subsubsection{HKUST}
Table~\ref{hk} lists the experimental results of the mentioned attention methods on HKUST task. We used the Chinese characters as the output units and character error rate (CER) to measure the ASR accuracy. 
We conducted more experiments thanks to the relatively small size of HKUST task.
The conclusions got on HKUST task were similar to those on LibriSpeech task: 1. sMoChA outperformed HMA and MoChA; 2. higher chunk order decoding methods always performed better; 3. the performance of MTA is best among the mentioned streaming attention methods, and is the same as that of LoAA. \par

Taking the results of both LibriSpeech and HKUST tasks into consideration, we can make the conclusion that our proposed streaming MTA method will not cause ASR accuracy degradation and is robust across different languages. \par

\begin{table}
    \centering
    \caption{Comparison of T-CTC and CTC Prefix Scores on LibriSpeech (WER) and HKUST (CER). When using T-CTC prefix score, we used the DWJD algorithm by default.}
    \scriptsize
    \begin{tabular}{c| c| c| c| c| c}
    \Xhline{2\arrayrulewidth}
    Attention & 
    T-CTC & 
    \multicolumn{2}{c|}{LibriSpeech} &
    \multicolumn{2}{c}{HKUST} \\

    Method & 
    Prefix & 
    \multicolumn{2}{c|}{test-clean / other} &
    \multicolumn{2}{c}{dev / eval} \\    
    \cline{3-6}
     &
      Score&
     VBL &
     VBS &
     VBL &
     VBS \\
    \hline
    LoAA &
    $\times$ & 
    3.9\,/\,12.2 &
    4.4\,/\,14.2 &
    28.7\,/\,27.6 &
    29.5\,/\,28.2 \\
    \hline
    \multirow{2}{*}{MTA}  & 
    $\times$ & 
    4.0\,/\,12.2 &
    4.8\,/\,14.2 &
    28.5\,/\,27.6 &
    29.4\,/\,28.1\\
    \cline{2-6}
      &
    $\surd$ &
    4.0\,/\,12.2 &
    4.8\,/\,14.2 &
    28.6\,/\,27.7 &
    29.8\,/\,28.3 \\
    \Xhline{2\arrayrulewidth}
    \end{tabular}
    \label{ctc}
\end{table}

\begin{table*}
    \centering
    \caption{Comparison of online end-to-end models with different encoder types.}
    \scriptsize
    \begin{tabular}{c|c|c|c|c|c|c}
    \Xhline{2\arrayrulewidth}
    \multirow{3}{*}{Encoder Type} &
    \multirow{3}{*}{Hop Frames} & 
    \multirow{3}{*}{Future Frames} & 
    \multirow{3}{*}{Decoding Method} &
    \multirow{3}{*}{Online} &
    LibriSpeech &
    HKUST \\
     &
     &
     &
     &
     &
    test-clean / other &
    dev / eval \\
    \cline{6-7}
     &
     &
     &
     &
     &
    WER (\%) &
    CER (\%) \\
    \hline
    VBL (Baseline) &
    -- &
    -- &
    LoAA / CTC Joint Decoding &
    No &
     3.9 / 12.2 &
     28.7 / 27.6\\
    \hline
    \multirow{2}{*}{VUL} &
    \multirow{2}{*}{--} &
    \multirow{2}{*}{--} &
    CTC Decoding &
    \multirow{5}{*}{Yes} &
    7.2 / 22.1 &
    34.6 / 32.5 \\
    \cline{4-4}\cline{6-7}
     &
     &
     &
    MTA / T-CTC Joint Decoding &
     &
    5.9 / 18.8 &
    33.6 / 31.7 \\
    \cline{1-4}\cline{6-7}
    \multirow{3}{*}{LC-VBL} &
    16 &
    16 &
    \multirow{3}{*}{MTA / T-CTC Joint Decoding} &
     &
    4.7 / 13.8 &
    30.1 / 28.6 \\
    \cline{2-3}\cline{6-7}
     &
    64 &
    16 &
     &
     &
    4.6 / 13.6 &
    30.2 / 28.4 \\
    \cline{2-3}\cline{6-7}
     &
    64 &
    32 &
     &
     &
    4.2 / 13.3 &
    29.4 / 27.8 \\  
    \Xhline{2\arrayrulewidth}
    VBS (Baseline) &
    -- &
    -- &
    LoAA / CTC Joint Decoding &
    No &
     4.4 / 14.2 &
     29.5 / 28.2\\
    \hline
    \multirow{2}{*}{VUS} &
    \multirow{2}{*}{--} &
    \multirow{2}{*}{--} &
    CTC Decoding &
    \multirow{5}{*}{Yes} &
    7.5 / 22.5 &
    35.7 / 33.2 \\
    \cline{4-4}\cline{6-7}
     &
     &
     &
    MTA / T-CTC Joint Decoding &
     &
    6.1 / 19.0 &
    34.8 / 32.2 \\
    \cline{1-4}\cline{6-7}
    \multirow{3}{*}{LC-VBS} &
    16 &
    16 &
    \multirow{3}{*}{MTA / T-CTC Joint Decoding} &
     &
    5.7 / 16.1 &
    30.8 / 29.1 \\
    \cline{2-3}\cline{6-7}
     &
    64 &
    16 &
     &
     &
    5.6 / 16.0 &
    30.6 / 29.2 \\
    \cline{2-3}\cline{6-7}
     &
    64 &
    32 &
     &
     &
    5.3 / 14.9 &
    30.4 / 28.9 \\     
    \Xhline{2\arrayrulewidth}
    \end{tabular}
    \label{online}
\end{table*}

\subsection{Results of T-CTC Prefix Score}
In this section, we examined how well the T-CTC prefix score approximated the offline CTC prefix score. Table \ref{ctc} lists the results of T-CTC prefix score on both LibriSpeech and HKUST tasks. We used PASM units on LibriSpeech task and Chinese character on HKUST task. The first line shows the results for the offline baselines. The second line shows the results of the MTA based models with CTC prefix score. 
For the models in the third line, we used MTA to stream attention, T-CTC to stream the CTC prefix score calculation and DWJD algorithm to stream the joint CTC/attention decoding. From table \ref{ctc}, we can see that, with all the above streaming methods applied, the proposed method only caused very slight degradation in ASR accuracy. \par

\subsection{Results of Low-latency Encoder}
In this section, we applied low-latency encoders, MTA, T-CTC prefix scores and DWJD algorithm to our online CTC/attention end-to-end models. We chose the Uni-LSTM and LC-BLSTM as low-latency encoders for comparisons. Uni-LSTM only needs history information while LC-BLSTM exploits finite future information. LC-BLSTM performs on the segmented frame chunks. Each chunk contains $N_{c}$ current frames and $N_{r}$ future frames, and LC-BLSTM hops $N_{c}$ frames at each time-step. The latency of the LC-BLSTM is limited to $N_{r}$ \cite{zhang2016highway}. 
It should be noted that we segmented the input frames according to $N_{c}$ and $N_{r}$, and the padding of VGG-style CNN-blocks would not bring about extra latency. \par

The results of different encoders are listed in table~\ref{online}. First, for the comparison in lines 2-3, we found that the proposed online CTC/attention based models achieved consistent accuracy improvements over the simpler Uni-LSTM end-to-end models trained by CTC objective, which was similar to the findings in other end-to-end systems \cite{watanabe2017hybrid, qiujia2019}. Second, for the comparisons in lines 3-6, we can see that the LC-BLSTM based encoder is superior to the Uni-LSTM based encoder for end-to-end ASR systems, which was also reported in \cite{zhang2016highway,jain2019lcblstm}. Third, in the LC-BLSTM encoder, we got significant improvement by increasing $N_{r}$ but minor improvement by increasing the $N_{c}$. This is because the LC-BLSTM, which carries the historical information across chunks, can make use of the historical information no matter how long the $N_{c}$ is, but it only obtains future context from $N_{r}$ future frames. \par

Finally, the proposed online hybrid CTC/attention model with 320 ms latency obtained 4.2\% and 13.3\% WERs on the test-clean and test-other sets, respectively. 
In HKUST task, the online CTC/attention model obtained 29.4\% and 27.8\% CERs on the development and evaluation sets, respectively. Compared with the offline baselines, our online end-to-end models exhibited acceptable performance degradation. More importantly, our online models only need limited future context, which significantly reduce the latency from utterance level to frame level and thus improve the user experience in human-computer interaction. \par

\subsection{Results of Decoding Speed}
Decoding speed is one concern when people deploy the online ASR systems. In this section, we evaluated the proposed online hybrid CTC/attention system, where the decoding was performed with beam sizes 1, 3, 5, 10 and 20.
We used VBL and LC-VBL encoders, and chose 64 and 32 for $N_{c}$ and $N_{r}$ for LC-VBL encoder, respectively.
The decoding time and CERs/WERs were measured by a computer with Intel(R) Xeon(R) Gold 6226 CPU, 2.70 GHz.
To further improve the decoding speed, we quantized the neural network activations and weights to 16-bits integers and the biases to 32-bits integers \cite{quantize2011}. \par

\begin{figure}
    \centering
    \includegraphics[scale=0.5]{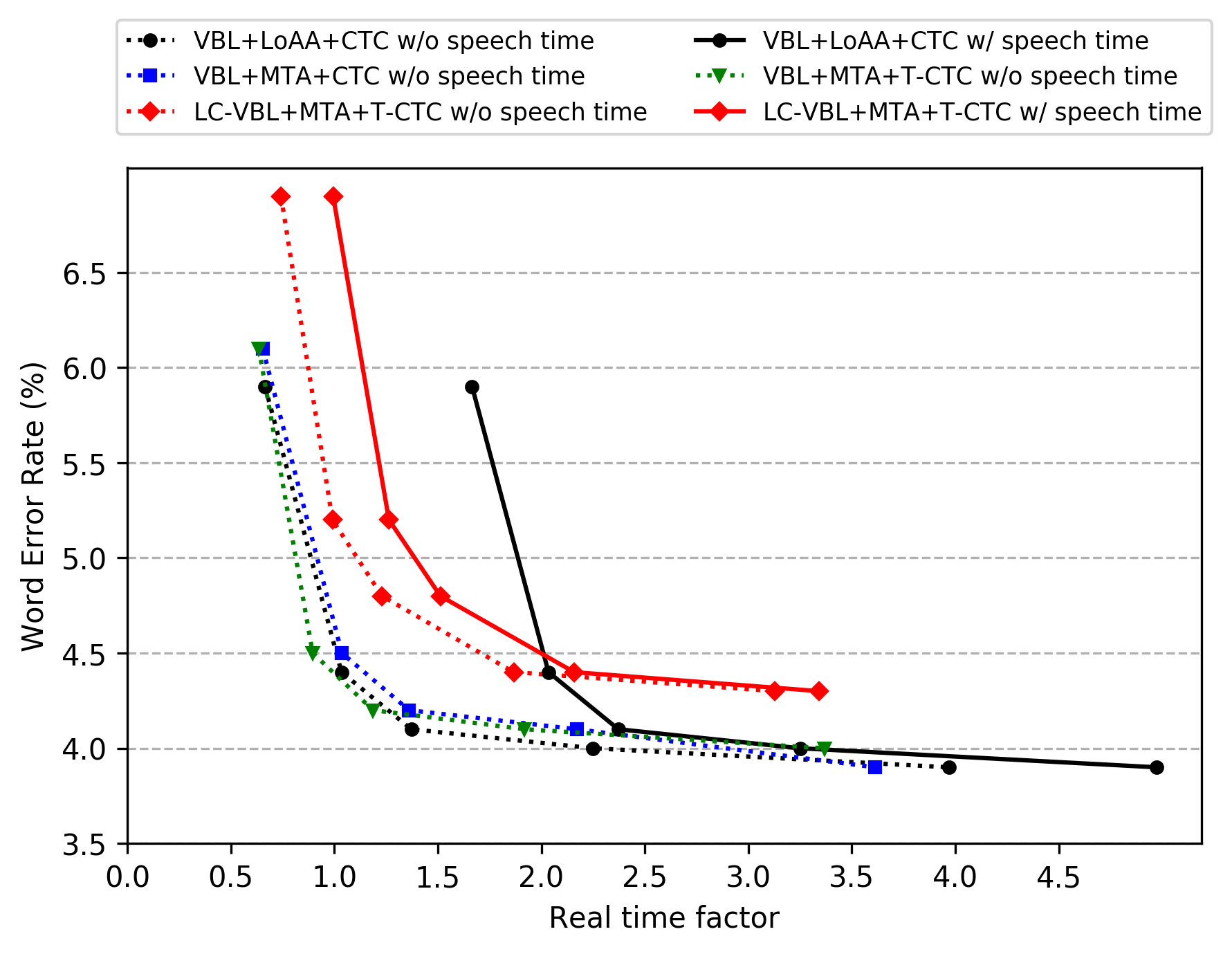}
    \caption{Real time factor (RTF) versus WERs for baseline and proposed methods on test-clean set of LibriSpeech task.}
    \label{fig:ls_speed}
\end{figure}

\begin{figure}
    \centering
    \includegraphics[scale=0.5]{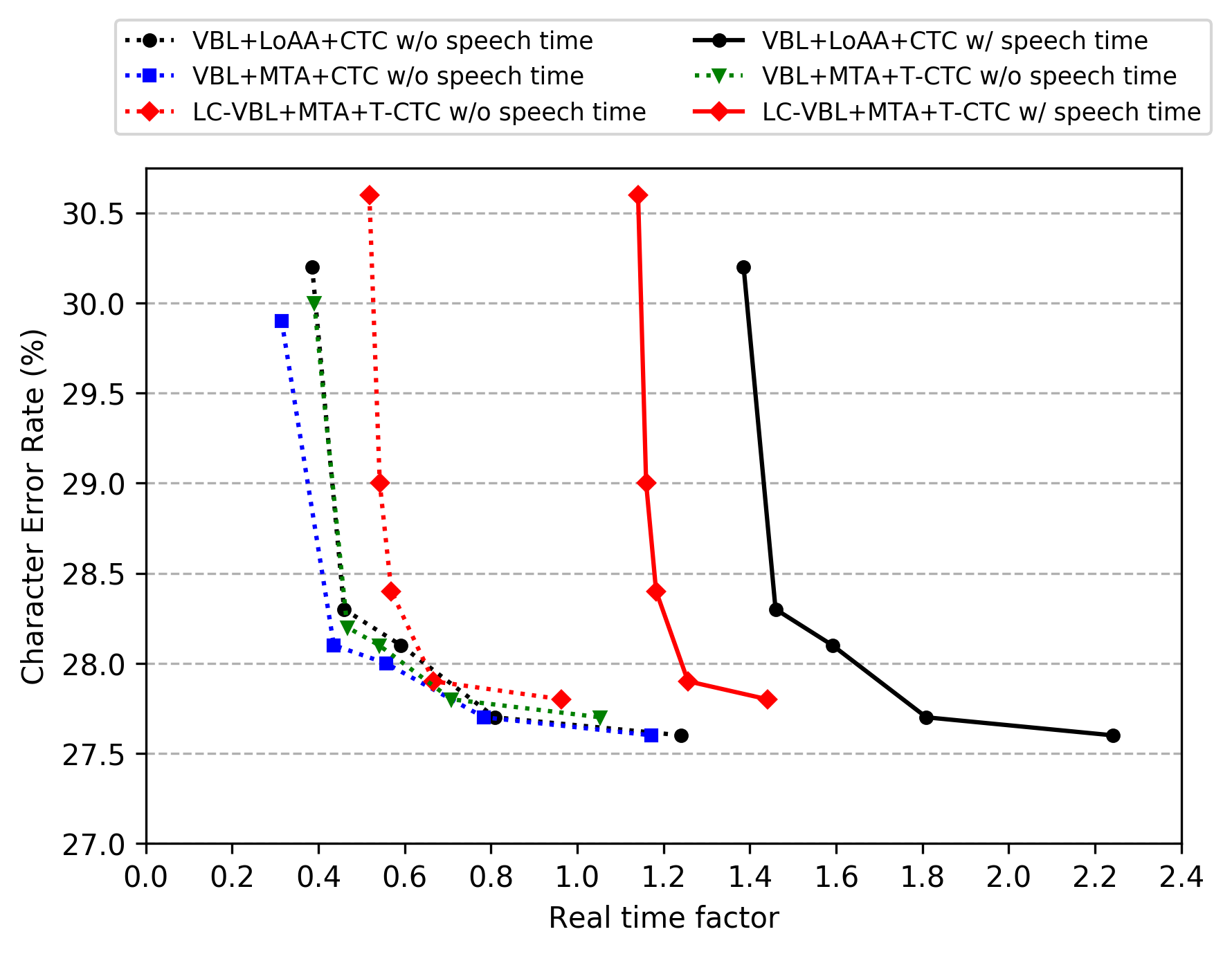}
    \caption{RTF versus CERs for models for baseline and proposed methods on dev set of HKUST}
    \label{fig:hk_speed}
\end{figure}

Figures~\ref{fig:ls_speed} and \ref{fig:hk_speed} display the relationships between the WERs/CERs and the real time factor (RTF) on LibriSpeech and HKUST, respectively. 
Comparing the black, blue and green dash lines in figures~\ref{fig:ls_speed} and \ref{fig:hk_speed}, MTA and T-CTC prefix scores indeed accelerated the decoding speed, especially when the beam size was larger than 5. This is because MTA and T-CTC prefix scores have less computational cost. \par

Furthermore, in online human-computer interaction scenarios, we have to take the speech time into consideration when evaluating the real decoding time. To simulate speaking, we sent 10 audio frames to the ASR systems every 100 ms. Under this scenario, the offline ASR system will be suspended until all the audio frames have been received while the online one can process the speech as it receives audio frames. 
From figures~\ref{fig:ls_speed} and \ref{fig:hk_speed} we can see that the online systems (red solid lines) were consistently faster than the offline baselines (black solid lines).
Even though the model sizes of online and offline systems were almost the same, the online systems were around 1.5x faster.
The decoding acceleration is due to that our online systems can process speech as the user begins speaking, which not only reduce the decoding latency but also improve the CPU utilization. \par


\section{Conclusion}
This paper proposes an online hybrid CTC/attention end-to-end ASR architecture, which can decode the speech in the low-latency and real-time manner.
This architecture can be trained from the scratch and in an end-to-end way.
Compared with the simpler Uni-LSTM CTC online end-to-end ASR systems, the proposed online CTC/attention based systems achieved consistent accuracy improvements.
Compared with the offline CTC/attention baselines, the proposed online systems achieved comparable performance.
Furthermore, the proposed online systems have the advantages over the offline baselines in both decoding latency and decoding speed. 
In the decoding latency aspect, the proposed online systems require limited future audio frames while the offline baselines needs the full audio.
In the decoding speed aspect, the proposed online systems can recognize from streaming audio or as the user begins speaking, which improves the CPU utilization, and thus accelerates the decoding speed by around 1.5 times compared with its offline counterparts. \par

In the future work, we will explore more parallelizable neural network architectures because it is difficult to parallelize the training and decoding processes of recurrent neural network based architectures.
Besides, our experiments revealed that the proposed MTA, T-CTC prefix score and DWJD algorithms almost did not degrade the recognition accuracy. However the application of the low-latency encoders, which were LSTM and LC-BLSTM in this paper, dragged down the recognition accuracy.
Another future work is to adopt teacher-student learning approaches to reduce the latency and maintain the recognition accuracy at the same time. \par


%



\section*{Acknowledgment}
This work is supported by the National Natural Science Foundation of China (Nos. 11590774,11590772,11590770)

\ifCLASSOPTIONcaptionsoff
  \newpage
\fi



\bibliographystyle{IEEEtran}
\bibliography{./IEEEexample}
%

%

%
\begin{IEEEbiography}[{\includegraphics[width=1in,height=1.25in,clip,keepaspectratio]{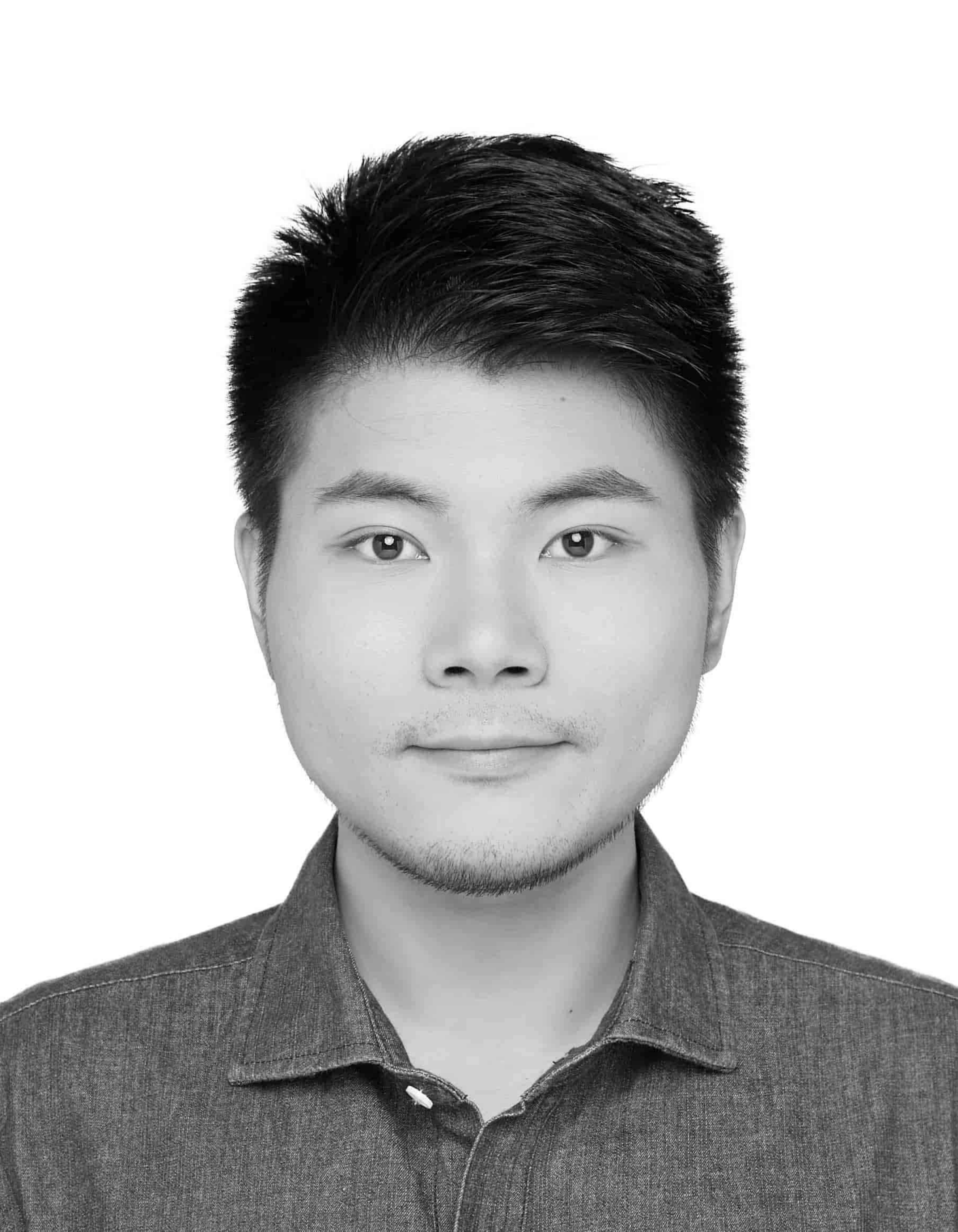}}]{Haoran Miao}received the B.S. degree in physics from the Nanjing University, Nanjing, China, in 2017. He is currently working toward the Ph.D. degree at the Institute of Acoustics, Chinese Academy of Sciences, Beijing, China. His research interests include automatic speech recognition and deep learning.
\end{IEEEbiography}

\begin{IEEEbiography}[{\includegraphics[width=1in,height=1.25in,clip,keepaspectratio]{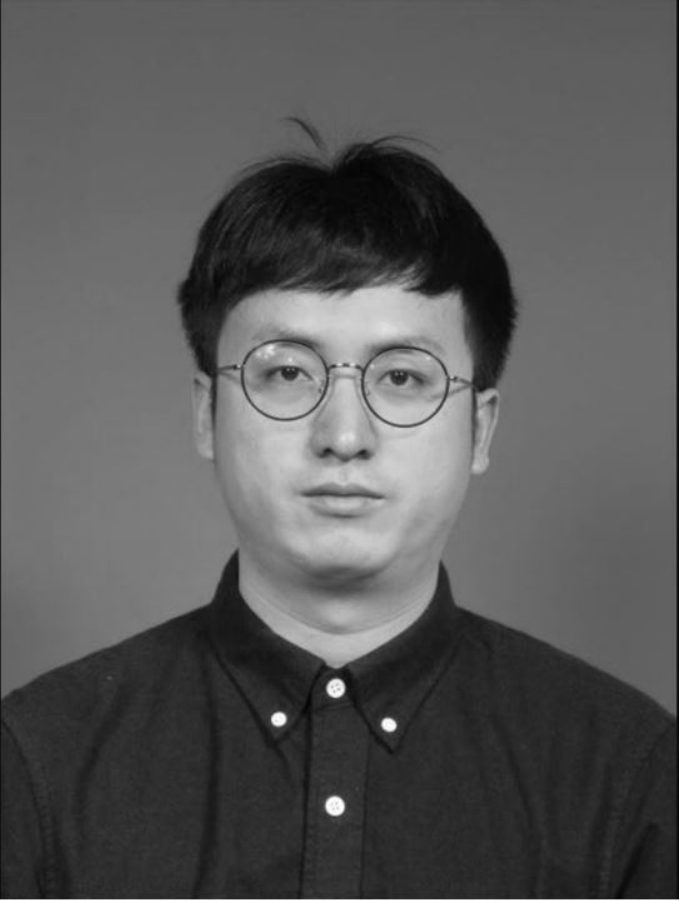}}]{Gaofeng Cheng}
received the B.S. degree in applied physics from the Beijing University of Posts and Telecommunications, Beijing, China, in 2014, and the Ph.D. degree in information and signal processing from the University of Chinese Academy of Sciences, Beijing, China, in 2019. 
He is currently an Assistant Professor with the Speech Acoustics and Content Understanding Laboratory, Chinese Academy of Sciences. His research
interests include speech recognition and deep learning.
\end{IEEEbiography}

\begin{IEEEbiography}[{\includegraphics[width=1in,height=1.25in,clip,keepaspectratio]{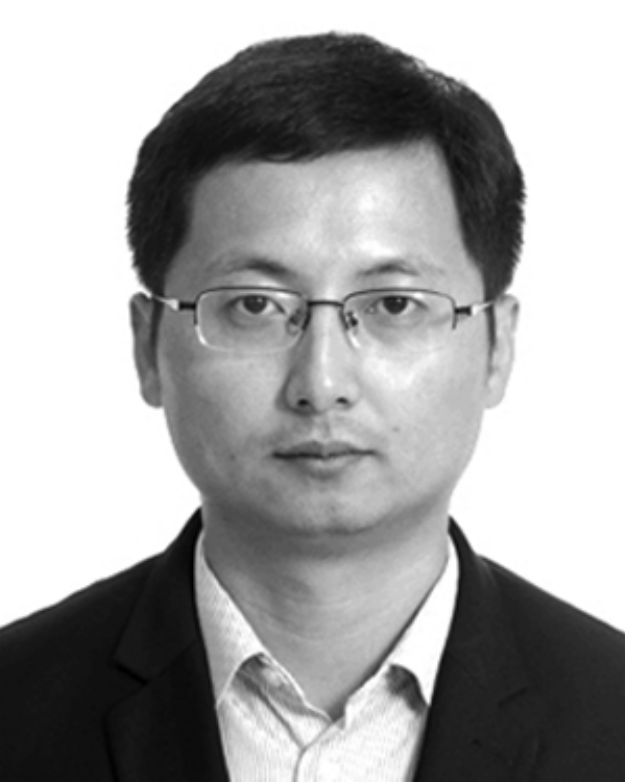}}]{Pengyuan Zhang}
received the Ph.D. degree in information and signal processing from the Institute of Acoustics, Chinese Academy of Sciences, Beijing, China, in 2007. From 2013 to 2014, he was a Research Scholar with the University of Sheffield. He is currently a Professor with the Speech Acoustics and Content Understanding Laboratory, Chinese Academy of Sciences. His research interests include spontaneous speech recognition, speech synthesis, and acoustic signal detection.
\end{IEEEbiography}
\begin{IEEEbiography}[{\includegraphics[width=1in,height=1.25in,clip,keepaspectratio]{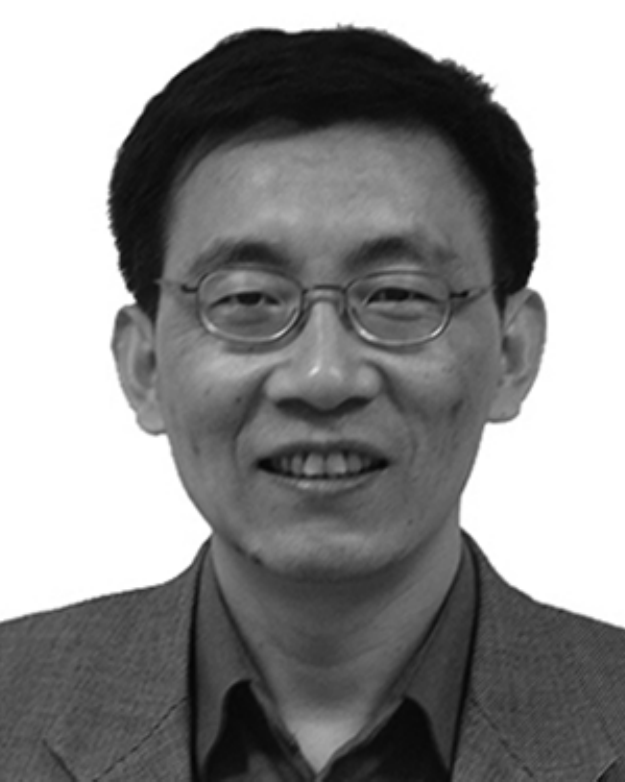}}]{Yonghong Yan}
received the B.E. degree in electronic engineering from Tsinghua University, Beijing, China, in 1990, and the Ph.D. degree in computer science and engineering from the Oregon Graduate Institute of Science and Technology, Hillsboro, OR, USA, in 1995. He is currently a Professor with the Speech Acoustics and Content Understanding Laboratory, Chinese Academy of Sciences, Beijing, China. His research interests include speech processing and recognition, language/speaker recognition, and human computer interface.
\end{IEEEbiography}






\end{document}